\documentclass{aa}  
\usepackage{graphicx, rotating}
\usepackage{aalongtable}
\usepackage{txfonts}
\newcommand{\mincir}{\raise -2.truept\hbox{\rlap{\hbox{$\sim$}}\raise5.truept
\hbox{$<$}\ }}
\newcommand{\magcir}{\raise -2.truept\hbox{\rlap{\hbox{$\sim$}}\raise5.truept
\hbox{$>$}\ }}
\newcommand{\siml}{\raise -2.truept\hbox{\rlap{\hbox{$\sim$}}\raise5.truept
\hbox{$<$}\ }}
\newcommand{\simg}{\raise -2.truept\hbox{\rlap{\hbox{$\sim$}}\raise5.truept
\hbox{$>$}\ }}
\newcommand{\be}{\begin{equation}}
\newcommand{\ee}{\end{equation}}
\newcommand{\ba}{\begin{eqnarray}}
\newcommand{\ea}{\end{eqnarray}}
\newcommand {\kpc} {$h_{70}^{-1}$ kpc $\;$}

\newcommand {\h} {$h_{70}^{-1}$ Mpc$\;$}
\newcommand {\hh} {$h_{70}^{-1}$ Mpc}
\newcommand {\hhh} {\;h_{70}^{-1} \mathrm{Mpc}}
\newcommand {\ks} {km~s$^{-1} \;$}
\newcommand {\kss} {km~s$^{-1}$}

\newcommand {\mquaa} {$\times 10^{14}\;h_{70}^{-1}\;\rm{M_{\odot}}$}
\newcommand {\mquaaa} {\times 10^{14}\;h_{70}^{-1}\;\rm{M_{\odot}}}
\newcommand {\mqui} {$\times 10^{15}\;h_{70}^{-1}\;\rm{M_{\odot}} \;$}
\newcommand {\mquii} {$\times 10^{15}\;h_{70}^{-1}\;\rm{M_{\odot}}$}
\newcommand {\ml} {$h_{70}\;\rm{M_{\odot}/L_{\odot}} \;$}

\newcommand{\arcm}{\ensuremath{\mathrm{^\prime}\;}}

\newcommand{\ldod} {$\times 10^{12}\;h_{70}^{-2}\;\rm{L_{\odot}}\;$}

\newcommand {\lumx} {$h_{70}^{-2} \;$}

\begin{document}
   \title{Fossil Groups Origins: I. RX J105453.3+552102 a very massive and relaxed system at z$\sim$0.5}

   \author{J. A. L. Aguerri\inst{1,2} \and M. Girardi\inst{3,4} \and W. Boschin\inst{5} \and R. Barrena\inst{1,2} \and J. M\'endez-Abreu\inst{1,2} \and R. S\'anchez-Janssen\inst{1,6} \and S. Borgani\inst{3,4} \and N. Castro-Rodriguez\inst{1,2} \and E. M. Corsini\inst{7} \and C. del Burgo\inst{8} \and E. D'Onghia\inst{9} \and J. Iglesias-P\'aramo\inst{10,11} \and N. Napolitano\inst{12} \and J. M. Vilchez\inst{10}
}

   \offprints{jalfonso@iac.es}

   \institute{Instituto de Astrof\'{\i}sica de Canarias. C/ V\'{\i}a L\'actea s/n, 38200 La Laguna, Spain \and Departamento de Astrof\'{\i}sica, Universidad de La Laguna. C/ Astrof\'{\i}sico Francisco S\'anchez, 38200 La Laguna, Spain. 
\and Dipartimento di Fisica-Sezione Astronomia of the Universit\`a degli Studi di Trieste, via Tiepolo 11, 34143 Trieste, Italy 
\and INAF - Osservatorio Astronomico di Trieste, via Tiepolo 11, 34143 Trieste, Italy\and Fundaci\'on Galileo Galilei-INAF, Rambla Jos\'e Ana Fern\'andez P\'erez 7, 38712 Bre\~na Baja, La Palma, Spain \and European Sourthem Observatory, Alonso de C\'ordova 3107, Vitacura, Santiago, Chile \and Dipartimento di Astronomia, Universit\`a di Padova, vicolo dell'Osservatorio 3, 35122 Padova, Italy \and UNINOVA/CA3, Campus da FCT/UNL, Quinta da Torre, 2825-149 Caparica, Portugal \and Harvard-Smithsonian Center for Astrophysics, 60 Garden Street, Cambridge, MA 02138, USA \and Instituto de Astrof\'{\i}sica de Andalucia--C.S.I.C., E-18008 Granada, Spain \and Centro Astron\'omico Hispano Alem\'an, C/ Jes\'us Durb\'an Rem\'on 2-2. 04004, Almer\'{\i}a, Spain \and INAF-Osservatorio Astronomico di Capodimonte, Salita Moiariello, 16, 80131, Napoli, Italy 
}

\date{Received  / Accepted }

\abstract {The most accepted scenario for the origin of fossil groups
  is that they are galaxy associations in which the merging rate was
  fast and efficient. These systems have assembled half of their mass
  at early epoch of the Universe, subsequently growing by minor
  mergers, and therefore could contain a fossil record of the galaxy structure
  formation.}{We have started an observational project in order to
  characterize a large sample of fossil groups. In this paper we present the
  analysis of the fossil system RX J105453.3+552102.}{Optical deep
  images were used for studying the properties of the brightest group
  galaxy and for computing the photometric luminosity function of the
  group. We have also performed a detail dynamical analysis of the
  system based on redshift data for 116 galaxies. Combining galaxy
  velocities and positions we selected 78 group members.}{RX
  J105453.3+552102 is located at $\left<z\right>=0.47$, and 
      shows a quite large line--of--sight velocity dispersion
  $\sigma_{v}\sim 1000$ \kss. Assuming the dynamical equilibrium,
  we estimated a virial mass of $M(<R_{\rm 200}) \sim 1$ \mqui.  No
  evidence of substructure was found within 1.4 Mpc
      radius. Nevertheless, we found a statistically significant departure from Gaussianity of the group members velocities in the most external regions of the group. This could indicate the presence of galaxies in radial orbits in the external region of the group. We also found that the photometrical luminosity function is bimodal, showing a lack of
  $M_{r}\sim-19.5$ galaxies. The brightest group galaxy shows low S\`ersic parameter ($n\sim2$) and a small peculiar velocity. Indeed, our accurate photometry
      shows that the difference between the brightest and the second
      brightest galaxies is 1.9  mag in the $r$-band, while the classical
      definition of fossil group is based on a magnitude gap of 2.}
          {We conclude that RX J105453.3+552102 does not follow the
            empirical definition of fossil group. Nevertheless, it is
            a massive, old and undisturbed galaxy system with little
            infall of $L^{*}$ galaxies since its initial collapse.}


   \maketitle
%

\section{Introduction}
\label{intr}

According to current cosmological cold dark matter (CDM) theories, structure in the Universe is built up hierarchically. Thus, virialized CDM halos grow by the merging of smaller virialized halos. Galaxies that populate these halos, also grow hierarchically by the merging of pre-existing galaxies (e.g., White \& Rees \cite{white78}). The efficiency of this hierarchical formation depends on several properties of the host dark matter (DM) halo and its satellites, among others: the galaxy number density, galaxy group kinematics, or formation epoch. Nevertheless, it may be possible to find galaxy associations in which the above properties conspire in a way that the merging was fast and efficient. These systems exist in the Universe, and are called fossil groups (FGs).  

The first identification of such a system was made by Ponman et al. (\cite{ponman94}), when they suggested that the elliptical galaxy-dominated system RX J1340.6+4018 was probably the relic of what previously constituted a group. Some years later, Jones et al. (\cite{jones03}) gave the first observational definition of FGs. Thus, from the observational point of view, FGs are characterized by the existence of an extreme magnitude gap ($\Delta m_{12}>2$ in the $R$-filter) between the two brightest members of the system within half of its virial radius. These galaxy associations also show an extended bright X-ray emission ($L_{\rm X}>10^{42}$ $h_{50}^{-2}$ erg s$^{-1}$) surrounding the brightest group galaxy. According to this definition, these systems are as common as poor and rich galaxy clusters together ($n\sim (1-4) \times 10^{-6}$ $h_{50}^{-3}$ Mpc$^{-3}$; Vikhlinin et al. \cite{vikhlinin98}; Jones et al. \cite{jones03}; Santos et al. \cite{santos07}; La Barbera et al. \cite{labarbera09}; Voevodkin et al. \cite{voevodkin10}).

Numerical simulations show that FGs could be particular cases of structure formation. Thus, according to these simulations, FGs have been formed inside highly concentrated DM halos at an early epoch of the Universe, assembling half of their dark matter mass at $z>1$, and subsequently growing by minor mergers. In contrast, non-fossil groups show, on average, a later formation (D'Onghia et al. \cite{donghia05}; von Benda-Beckmann et al. \cite{vonbenda08}). This early formation leaves enough time for $L^{*}$ galaxies to merge into a massive elliptical-type galaxy located at the center of the group, producing a lack of intermediate-luminosity galaxies and a large magnitude gap between the brightest and the second brightest galaxy of the group.  FGs also have special dynamical properties which speed up the merging efficiency. In particular, the in-fall of massive satellites in FGs took place on orbits with low angular momentum, which might be the main responsible of the anisotropy of the group galaxies, in such a way that groups with highly radially anisotropic velocity distributions tend to become fossil (Sommer-Larsen \cite{somer06}). Simulations also indicate that FGs have only been able to accrete on average one galaxy since $z\sim 1$, compared to $\sim 3$ galaxies for normal groups (see von Benda-Beckmann et al. \cite{vonbenda08}). This means that FGs provide unique clues on the history of cosmic mass assembly and the relationship between baryons and their host halos. They also could 
have a fossil record of the structure formation of galaxies at early epochs of the Universe.

Observations are broadly in agreement with the formation framework of FGs proposed by numerical simulations. Thus, Khosroshahi et al. (\cite{khosro07}) compared the scaling relations of a sample of FGs and non-fossil systems and found that FGs follow the X-ray luminosity-temperature relation ($L_{\rm X}$-$T_{\rm X}$) as clusters and groups. However, there are significant differences in the optical vs. X-ray luminosities ($L_{\rm opt}-$$L_{\rm X}$), X-ray luminosity vs. cluster velocity dispersion ($L_{\rm X}-\sigma$) and X-ray temperature vs. cluster velocity dispersion ($T_{\rm X}-\sigma$) relations. In particular, for a given $\sigma$, FGs are located in more luminous and hotter X-ray halos than normal groups and clusters. They also have larger X-ray luminosities than normal groups for a given $L_{\rm opt}$ (but see also Voevodkin et al. \cite{voevodkin10}). These differences could be due to an early formation epoch of FGs as suggested by simulations (Khosroshahi et al. \cite{khosro07}). Detailed X-ray observations of some FGs also indicate that these systems were assembled at early epochs in high centrally concentrated DM halos with large mass-to-light-ratio ($M/L$) relations. Nevertheless, they do not show cooling cores as those detected in galaxy clusters, which points toward the presence of other heating mechanisms, like AGN feedback (Sun et al. \cite{sun04}; Khosroshahi et al. \cite{khosro04},\cite{khosro06}; Mendes de Oliveira et al. \cite{mendes09}). The absence of recent galaxy or cluster major mergers together with the lack of cool cores make FGs the ideal objects to study the effects of AGN feedback and the link between galaxy evolution and intra-group medium (IGM).

Optical and near-infrared observations indicate that the faint-end slope ($\alpha$) of the luminosity function (LF) of FGs spans a wide range of values. Thus, the Schechter function fitted to the LF of these systems shows values in the range -1.6$<\alpha<$-0.6 (Cypriano et al. \cite{cypriano06}; Khosroshahi et al. \cite{khosro06}; Mendes de Oliveira et al. \cite{mendes06}, \cite{mendes09}). This suggests that some FGs are dwarf rich systems like  similar size/mass galaxy clusters, while others show a lack of dwarf galaxies. It has been pointed out that these differences between fossil and non-fossil systems can reflect different substructure distribution (Jones et al. \cite{jones00}). Thus, FGs could have one order of magnitude less substructure with respect to the standard cosmological model predictions (D'Onghia \& Lake \cite{donghia04}). Nevertheless, the number of LFs measured for FGs is scarce and the system where this different substructure was measured has only 40$\%$ of the Virgo mass. It should be pointed out the fact that many systems classified in the past as FGs turned to be fossil clusters.  On mass scale of groups it is not completely clear when the transition from galaxy formation to galaxy cluster formation happens. The low mass FGs are intermediate systems in this respect and can give hints of how and at which extent the substructures are accreted. Thus, studies on low mass FGs might give a hint on the abundance of dwarf galaxies in systems with mass scale intermediate between a galaxy and a galaxy cluster as compared to the standard cosmological predictions.

The brightest group galaxies (BGGs) located at the center of FGs are among the most massive galaxies known in the Universe. They contain the key for understanding the formation and evolution of FGs. Observations show that BGGs have also different observational properties than other bright elliptical (E) galaxies. In particular, they present discy isophotes in the center and their luminosity correlates with the velocity dispersion of the group (Khosroshahi et al. \cite{khosro06}). These different properties suggest a different formation scenario for bright Es in fossil and non-fossil systems. While bright Es in FGs would grow by gas-rich mergers, giant Es in non-fossil systems would suffer more dry mergers. However, recent samples of BGGs do not find these differences (La Barbera et al. \cite{labarbera09}). 

All previous results have the drawback that they were obtained using small samples of FGs. This could be the reason of some contradictory results found by different studies. The lack of a large and homogeneous statistical study of this kind of systems make the previous results not conclusive. A systematic study of a large sample of FGs remains to be done.

\subsection{Fossil Groups Origins (FOGO) project.}

We have started a large observing program on FGs. The aim of this project is to carry out a systematic, multiwavelength study of a sample of 34 FGs selected from the Sloan Digital Sky Survey (SDSS; Santos et al. \cite{santos07}). This sample is ideal for providing strong constraints on the observational properties of the galaxy populations in FGs due to its unique characteristics. The sample spans the last 5 Gyr of galaxy evolution ($0<z<0.5$). The groups have a large range of masses and therefore X-ray luminosities (0.04-30 $\times 10^{43}$ erg s$^{-1}$). This will be useful for the study of the dependence of FG properties on group mass. Indeed, the absolute magnitude of the central BGGs spans a large range of magnitudes ($-25.3<M_{r}<-21.25$), which will also allow us to analyze the relation of the BGGs and the cluster environment.

The specific scientific goals of this programme are: 

\begin{itemize}
\item Mass and dynamics of FGs: Given the early assembly of DM halos in FGs, it is reasonable to assume that they are more relaxed systems than non-FGs. Simulations show that using the kinematics of satellite galaxies it is possible to infer their dynamical status and mass distribution all the way to the virial radius. Moreover, the in-fall of $L^{*}$ galaxies occurs along filaments with small impact parameters, increasing the group concentration and therefore its merging efficiency. The signature of this filamentary in-fall could be reflected in the orbital anisotropy of satellite galaxies, which are expected to show an excess of radial anisotropy in their outer regions with respect normal groups (see e.g. Sommer-Larsen 2006). A thorough mass distribution and dynamical study of FGs like that performed by Biviano \& Katgert (\cite{biv04}) for galaxy clusters remains to be done.
\item Properties of galaxy populations in FGs: These systems are characterized by a large magnitude gap between the two brightest group members. The high luminosity of the BGG combined with the large magnitude gap hint at the merging of the most massive group members. However, little is known about the fate of low-mass satellites in FGs. The faint-end of the LF of low-mass FGs will tell us whether these systems present a paucity of dwarfs (as in the Local Group) or are more similar to dwarf-rich galaxy clusters. 
\item Formation of the BGG: Understanding the formation of the BGGs is one of the main goals of the project. Simulations indicate that BGGs are the outcome of a process of intense merging between galaxies. Moreover, depending of the gas content of the precursor galaxies, BGGs can present different isophotal structure and AGN activity, given that mergers regulate the formation and fuelling of active nuclei (di Matteo et al. \cite{dimatteo05}). Additionally, the determination of ages, metallicities, $\alpha$-enhancements, and kinematics of their stellar populations will provide strong constraints on the formation history of these unique objects. 
\item Extended diffuse light: If BGGs are the outcome of intense past merging processes, they are expected to have diffuse and extended stellar haloes. Numerical simulations (Sommer-Larsen \cite{somer06}) show that this diffuse component may contribute up to 40$\%$ of the total $V$-band light of the group, but is only detected at very faint surface brightness ($\sim26.5$ mag arcsec$^{-2}$). This fraction is comparable to or larger than that observed in nearby galaxy clusters and normal groups (Zibetti et al. \cite{zibetti05}; Aguerri et al. \cite{aguerri05}, \cite{aguerri06}; Castro-Rodriguez et al. \cite{castror03}, \cite{castror09}). Therefore, the detection of this component is crucial for understanding the formation history of FGs, and to provide an accurate determination of their total baryonic content.
\item Connection between BGG and intragroup medium: Numerical simulations and some observational results show that FGs are old and relaxed systems. Therefore, FGs should provide ideal environments for the formation of cool cores as those found in some normal groups and galaxy systems. Nevertheless, so far these cool cores are either not observed in FGs or they are much smaller than expected (Khosroshahi et al. \cite{khosro04}, \cite{khosro07}; Sun et al. \cite{sun04}; Mendes de Oliveira et al. \cite{mendes09}). The absence of major recent galaxy or cluster mergers together with the absence of cool cores make FGs the ideal objects to study the effects of other heating mechanisms, like AGN heating. The AGN feedback would deposit metals and energy into the IGM central gas. The wind metal injection would make the central SN Ia/SN II ejecta of FGs different from that or normal groups and similar size clusters. This scenario will be tested for some of our groups.
\item Theory and numerical simulations: the lack of observational data on FGs has so far prevented the validation of most theoretical results. The present programme will provide invaluable information on the mass distribution of FGs, the orbital characteristics of their galaxy populations, the abundance of dwarf galaxies, the amount and distribution of diffuse light, the inner structure of BGGs, and the AGN feedback. This unprecedented data-set is expected to challenge current numerical simulations and serve as reference for future ones.
\end{itemize}

In order to reach the previous scientific objectives, the Fossil Group Origins (FOGO) project was approved as an International Time Program (ITP) at the Roque de los Muchachos Observatory (ORM) in La Palma, Spain. The multiwavelength observations have been  taken during the period 2008--2010. The assigned telescopes for the project were the 4m  William Herschel Telescope (WHT), the 3.5m  Telescopio Nazionale Galileo (TNG), the 2.5m  Isaac Newton Telescope (INT), and the 2.5m Nordic Optical Telescope (NOT). The set of observations include optical and near-IR imaging, multi-object spectroscopy, and integral field spectroscopy. 

All FGs from Santos et al. (2007) have been imaged through the sloan $r$-band with the Wide Field Camera (WFC) at INT and the Andalucia Faint Object Spectrograph and Camera (ALFOSC) at NOT. We plan to reach $\mu_{r}\sim 26$ mag arcsec$^{-2}$ with $S/N\sim 1$ per pixel. In addition, near-IR images in the $K$-band have been taken for the central regions of 17 groups using  The Long-slit Intermediate Resolution Infrared Spectrograph (LIRIS) at the WHT. The multi-object spectroscopy was obtained with the Wide-field Fibre Optic Spectrograph (WYFFOS) at the WHT and the Device Optimized for Low Resolution spectrograph (DOLORES) at the TNG. The target galaxies for the multi-object spectroscopy have been located within 1.4 Mpc radius around the BGG, and with $m_{r}<22$ mag. The selection was done taking into account the photometrical redshift information given by SDSS (see Sec. 2.2). The integral field spectroscopy was obtained with INTEGRAL/WYFFOS mounted at WHT. These observations  provide integral field spectroscopy of the central regions of the BGGs. 

In this paper we present the first results of the survey with a detailed analysis of the optical images and multi-object spectroscopy of one of the FGs: RX J105453.3+552102. This paper is a pilot program in order to show the capabilities of the FOGO project. The paper is organized as follow. The observations are shown in Section 2. The galaxy catalogues are given in Section 3. The photometric and spectroscopic LFs are shown in Section 4. The photometric properties of the brightest group galaxy are described in Section 5. The internal dynamic of the group is analyzed in Section 6. The discussion and conclusions are given in Section 7.

Unless otherwise stated, we give errors at the 68\% confidence level
(hereafter c.l.). Throughout this paper, we use $H_0=70$ km s$^{-1}$
Mpc$^{-1}$ in a flat cosmology with $\Omega_0=0.3$ and
$\Omega_{\Lambda}=0.7$. In the adopted cosmology, 1\arcm corresponds
to 353 \kpc at $z=0.47$, the redshift of the group (see Sec. 3.3).

\section{Observations and data reduction}

\subsection{Optical imaging}

Optical imaging of RX J105453.3+552102 was carried out at the 2.5m NOT telescope in March 2008. The data were taken under photometric conditions and a typical seeing of FWHM$\sim$1$''$ during the run. The observations were centered at the position of the BGG ($\alpha(J2000)= 10^{\rm{h}}54^{\rm{m}}52^{\rm{s}}; \delta(J2000)= +55^{\rm{o}}21'12.5'')$. We used ALFOSC in image mode, with the SDSS $r$-band mounted in the filter wheel. The CCD detector has a size of 2048$\times$2048 pixels, with a plate scale of 0.19 arcsec/pixel, or 5.9 kpc/arcsec at the distance of the group ($z=0.47$). This implies that we have mapped a radius of $\sim1.1$ Mpc around the BGG.

The data reduction was performed using standard IRAF\footnote{IRAF is distributed by the
 National Optical Astronomy Observatories, which are operated by the
 Association of Universities for Research in Astronomy, Inc., under
 cooperative agreement with the National Science Foundation.} routines. The bias was subtracted from the images using a master bias obtained by the combination of 10 bias taken at the beginning of the night. We also obtained sky flat images during the twilight. The sky flatfields  were combined creating a master flat to correct the images.  Some residual light appears in the flat-fielded images. In order to have the best possible flat-field correction of the images, we  did a aditional flat-field correction using a super-flat obtained by the combination of the scientific images. The images were observed following a dithering pattern, which turned to be not large enough  in order to remove the objects from the images after the combination of all scientific images. In order to remove the objects and create a super-flat we developed our own procedure. The procedure starts by computing the sky background level and its standard deviation ($\sigma_{\rm sky}$) from the biased and flat-fielded images. We adopted the mode of the image as the background level. In the second step we mask out from the scientific images all pixels with 1$\times \sigma_{\rm sky}$ over the sky level. Those pixels were substituted by the value of the sky computed locally. This local background was obtained by measuring the mode in a box of 400-pixels size centered at each of the masked pixels. After several trials, this box size resulted the most appropriate in order to mask the emission from the objects. The masked images were finally combined, resulting a super-flat which contains the large scale light pattern not corrected by the twilight flats. This super-flat was used for a second flat correction of the scientific images. The residual structures in the background of the images were successfully corrected when this super-flat was used.

The bias, flat-field corrected scientific images were astrometrized and register into a common spatial reference. These images were combined into a final $r$-band scientific image with a total exposure time of 2.5 hours and a seeing of 1$''$. Figure 1 shows the $r$-band image of the group and a zoom around the BGG.

The image was calibrated comparing the SDSS $r$-band magnitudes of the stars located in the field of view (FOV) of our image with our instrumental magnitudes. A simple zero-point offset with the SDSS magnitudes was computed in order to calibrate the image. The rms of the calibration turned to be 0.08 mag.

\subsection{Spectroscopic data}
\label{spec}

Multi--object spectroscopic observations of RX J105453.3+552102 were carried out at
 the TNG telescope in February 2008. We used DOLORES in the Multi Object Spectroscopic mode (MOS) with the LR--B
 Grism 1, yielding a dispersion of 187 \AA/mm. We used the new E2V CCD
 detector with a pixel size of 13.5 $\mu$m. The CCD is a matrix of
 $2048\times2048$ pixels. We observed five MOS masks for a
 total of 151 slits.  We acquired 4 exposures of 1800 s for three masks
 and 5 exposures of 1800 s for two masks. Wavelength calibration was
 performed using Helium--Argon lamps. Reduction of spectroscopic data
 was carried out with the IRAF package.

The target selection for the multi--object spectroscopy was based on the SDSS database. We downloaded a catalogue with all galaxies with $m_{r}<22$ mag which are located within a radius of 1.4 Mpc around the BGG at the distance of the group. In a second step, we selected the galaxy targets using the photometric redshift information provided by SDSS. We selected as possible targets  galaxies with 0.37$<z_{\rm phot}<0.57$. This range of photometrical redshift was chosen because the group, according to the spectroscopic redshift of the BGG provided by SDSS, is located at $z=0.47$ and the typical photometric redshift error from SDSS is about 0.1. Figure~\ref{colormag} shows the color-magnitude diagram of the galaxies located within 1.4 Mpc radius from the group center. The magnitudes in the $g$ and $i$ bands were obtained from the SDSS database. The figure also shows the targets selected for the multi--object spectroscopy.

 Radial velocities, $\rm v=cz$, of the selected galaxies were
 determined using the cross--correlation technique (Tonry \& Davis
 \cite{ton79}) implemented in the IRAF package RVSAO (developed at the
 Smithsonian Astrophysical Observatory Telescope Data Center).  Each
 spectrum was correlated against six templates for a variety of galaxy
 spectral types: E, S0, Sa, Sb, Sc, Ir (Kennicutt \cite{ken92}). The
 template producing the highest value of $\cal R$, i.e., the parameter
 given by RVSAO and related to the signal--to--noise ratio of the
 correlation peak, was chosen. Moreover, all the spectra and their best
 correlation functions were examined visually to verify the redshift
 determination. In 24 cases (see Table~\ref{catalogFG10}) the redshift
 of the galaxies was determined by the emission lines observed in the
 wavelength range of the spectra (EMSAO procedure).

\begin{figure*}
\centering 
\includegraphics[width=8cm,angle=90]{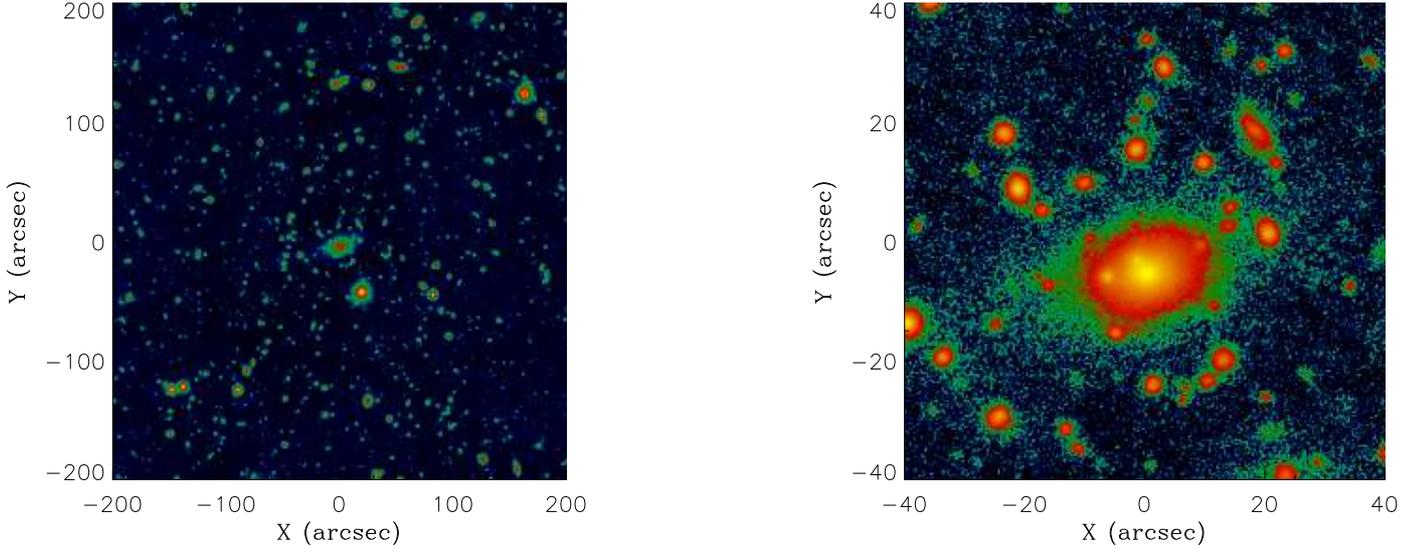}
\caption{(left panel) $r$-band images of RX J105453.3+552102; (right panel) zoom of the central region of the group. In both panels North is up and east is left.}
\label{image}
\end{figure*}

\section{Galaxy catalogues}

Two different catalogues were obtained. One catalogue contains the photometry of the galaxies located in our $r$-band image. This catalogue was used to compute the photometric luminosity function of the group. The other catalogue contains the velocity information of the galaxies observed in the MOS observations. This will be used for determining the group members, the spectroscopic luminosity function and the dynamics of the group.

\subsection{Photometric galaxy catalogue}

The photometric galaxy catalogue was obtained using SExtractor (Bertin \& Arnouts \cite{ber96}). SExtractor identifies objects and measures their flux in astronomical images. Here we discuss those parameters of SExtractor that are relevant for the identification and flux measurements of our objects.

Objects were identified as imposing that they cover a certain minimum area and have  number counts above a limiting threshold taking the sky local background as a reference. The limiting size and fluxes were 25 pixels and one standard deviation of the sky counts, respectively. The selected limiting size corresponds to an apparent size of 1 arcsec, which is about the size of the seeing disc. We have performed careful visual inspections of the frames in order to deal with the best combination of the above parameters that remove spurious objects from the catalogues. The resulting photometric catalogue contains 957 objects. For each object we measured three different magnitudes. Two of them were aperture magnitudes with diameters: 5.6 and 15 pixels. The smallest diameter corresponds to a circular aperture of area equal to the minimum detection area in SExtractor, and the largest one to an aperture of radius 3$\sigma$, being $\sigma$ the standard deviation of the Gaussian seeing point spread function (PSF) of the image. We have also included in the catalogue the $MAG-AUTO$ magnitude given by SExtractor. This magnitude is computed in a elliptical aperture enclosing the flux of each object.

The separation between galaxies and stars was performed on the basis of the SExtractor stellarity index ($S/G$). Objects with $S/G$ close to 1 correspond to stars while galaxies are those with $S/G$ close to 0. This separation is clear for bright objects. In contrast, a correct classification is more difficult for faint objects. In order to be conservative, we have determined stars as those objects with $S/G>$0.85.

We have determined the completeness of our photometric data following the same criteria as S\'anchez-Janssen et al. (\cite{sanchez05}). Thus, for all the detected objects we computed the mean central surface brightness ($\mu_{o}$) measured in a circular aperture of area equal to the minimum detection area used in SExtractor. Figure~\ref{comple} shows $\mu_{o}$ as a function of the apparent $r$-band magnitude ($m_{r}$) of the detected objects. We have overplotted with crosses those objects with $S/G>0.85$ corresponding to stars. Notice that stars (objects with $S/G>0.85$) are located in the upper diagonal of the plot. In contrast, the galaxies (objects with $S/G<0.85$) are located in the broader region of the plot. We have computed the central surface brightness which correspond to 1$\sigma_{\rm sky}$ detection limit used by SExtractor resulting $\mu_{o}=27.4$ mag arcsec$^{-2}$ (horizontal line in Fig.~\ref{comple}). Given the distribution of the objects in Fig.~\ref{comple} we can conclude that our photometric limiting magnitude is $m_{r}\sim24.0$ mag.

\begin{figure}
\centering 
\resizebox{\hsize}{!}{\includegraphics{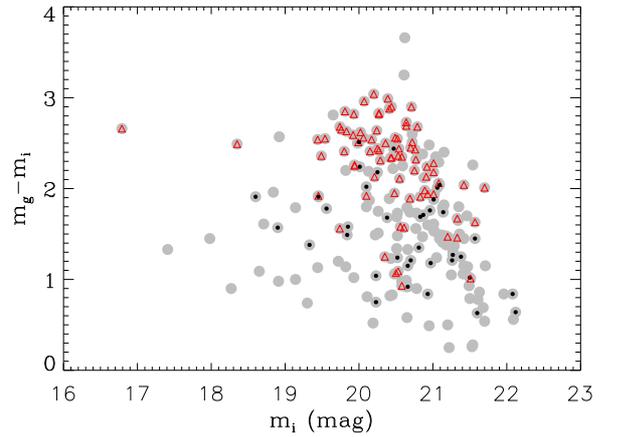}}
\caption{(Color-magnitude diagram of the SDSS galaxies located within 1.4 Mpc around the BGG (grey large circles). The black points (non cluster members) and red triangles (cluster members) represent the galaxies selected for the spectroscopy.}
\label{colormag}
\end{figure}

\begin{figure}
\centering 
\resizebox{\hsize}{!}{\includegraphics{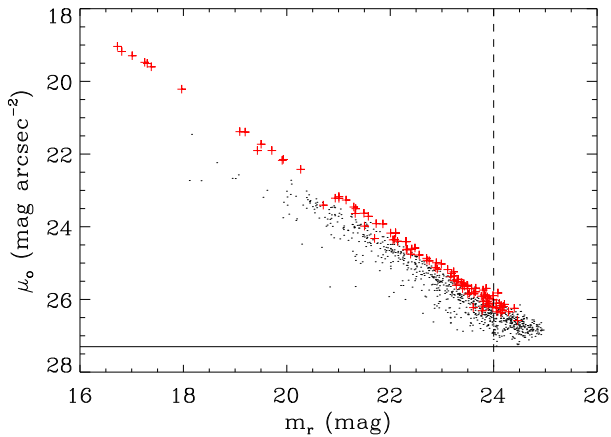}}
\caption
{The aperture central surface brightness versus apparent magnitude of the detected objects. Crosses represent stars (objects with $S/G>0.85$) and points correspond to galaxies (objects with $S/G<0.85$). The horizontal line correspond to 1$\sigma_{\rm sky}$ isophotal detection limit used in SExtractor. The vertical dashed line is our limiting magnitude.}
\label{comple}
\end{figure}

\subsection{Spectroscopic galaxy catalogue}

The spectroscopic catalogue was formed by the galaxies with measured
radial velocities from our MOS observations. Our spectroscopic survey
in the field of RX J105453.3+552102 consists of spectra for 116 single
galaxies, of which 18 (4) have double (triple) measurements as coming
from five different masks. The nominal errors as given by the
cross--correlation are known to be smaller than the true errors (e.g.,
Malumuth et al. \cite{mal92}; Bardelli et al. \cite{bar94}; Ellingson
\& Yee \cite{ell94}; Quintana et al. \cite{qui00}).  Double/triple
redshift determinations for the same galaxy allow to estimate more reliable errors. Previous analyses on data acquired with the
    same instrumentation and of comparable quality showed that nominal
    errors obtained through the RVSAO procedure should be multiplied
    by a factor $\sim 2$ (Barrena et al. \cite{bar09} and
    references therein). Following the method of Barrena et
    al. (\cite{bar09}), we compared the determinations coming from
    different masks and found that the nominal errors were
    underestimated by a factor 2 in our case, too. Therefore, 
    we assumed that the true errors are larger than nominal
    cross--correlation error by a factor of 2.  To check the errors
    obtained trough the EMSAO procedure we had only three galaxies
    with double determinations.  To be conservative we assumed that
    errors on $cz$ recovered from EMSAO are $\sim$ 100 \kss.

\begin{figure}
\centering
\resizebox{\hsize}{!}{\includegraphics{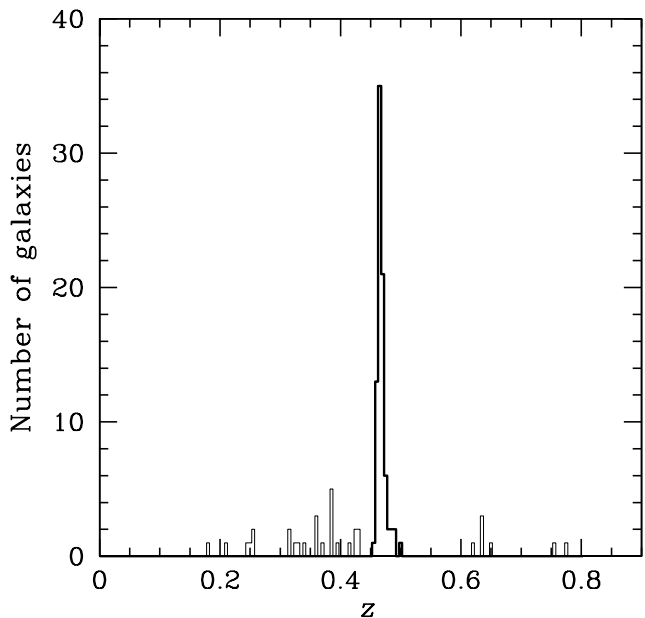}}
\caption
{Redshift galaxy distribution. The thick solid line histogram refers to the
  83 galaxies assigned to the RX J105453.3+552102 complex according to the DEDICA
  reconstruction method.}
\label{fighisto}
\end{figure}

As for the compilation of our spectroscopic catalogue, for all the
galaxies with multiple redshift estimates we used the weighted mean of
the multiple measurements and the corresponding errors.  The median
error on $cz$ was 98 \kss. The redshift distribution of the 116 galaxies having 
robust radial velocity measurements is shown in Fig.~\ref{fighisto}.

\begin{figure}
\centering 
\resizebox{\hsize}{!}{\includegraphics{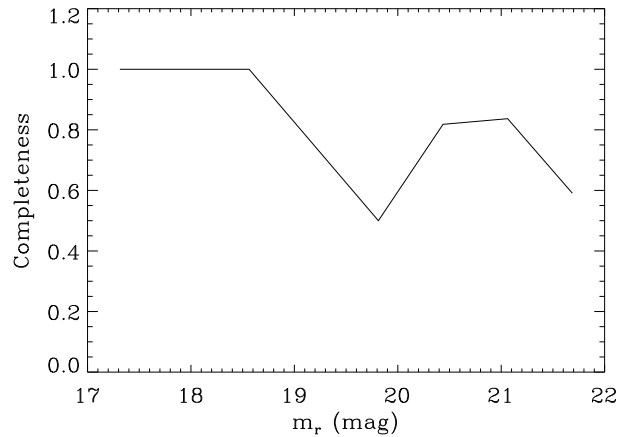}}
\caption{Completeness of MOS observations (see text for more details).}
\label{comple_lf}
\end{figure}

We measured redshifts
for galaxies down to magnitude m$_{r}\lesssim$ 22 mag. Nevertheless, the median completeness of the observations was 84$\%$, being $100\%$ only for galaxies with magnitude m$_{r} \lesssim 18.5$ mag (see Fig.~\ref{comple_lf}). This spectrocopic completeness will be taken into account for the computation of the spectroscopic LF of the group (see Sect. 4).

Table~\ref{catalogFG10} lists the photometric and kinematic properties
of the galaxies with MOS data. The columns indicate: (Col 1)
    galaxy name; (Col 2) our SExtractor $r$ band magnitude (the error of this magnitude is in the range 0.08-0.1 mag for all objects); (Col 3)
    line-of-sight velocity $v=cz$; (Col 4) 1=cluster member, 0=non
    cluster member; (Col 5) comments: E=redshift measured using
    emission lines; NE= galaxies with no envidence of
    obvious emission lines; (Col 6-10) $ugriz$ SDSS model magnitudes.

\subsection{Member selection}
\label{memb}

To select group members out of 116 galaxies having redshifts, we
followed a two steps procedure. First, we performed the 1D
adaptive--kernel method (hereafter DEDICA, Pisani \cite{pis93} and
\cite{pis96}; see also Fadda et al. \cite{fad96}; Girardi et
al. \cite{gir96}). We searched for significant peaks in the velocity
distribution at $>$99\% c.l.. This procedure detects RX J105453.3+552102 as a peak
at $z=0.47$ populated by 83 galaxies considered as candidate
group members (in the range $0.456<z<0.502$, see
Fig.~\ref{fighisto}). Out of 33 non members, 26 and 7 were
foreground and background galaxies, respectively. Notice the advantage of including the photometrical redshifts in the selection of the spectroscopical targets. Thus, 88$\%$ of the galaxies have  redshifts in the range 0.37$<z<$0.57, the range of photometric redshifts we selected. Indeed, 70$\%$ of all measured redshifts correspond to actual members.

\begin{figure}
\centering 
\resizebox{\hsize}{!}{\includegraphics{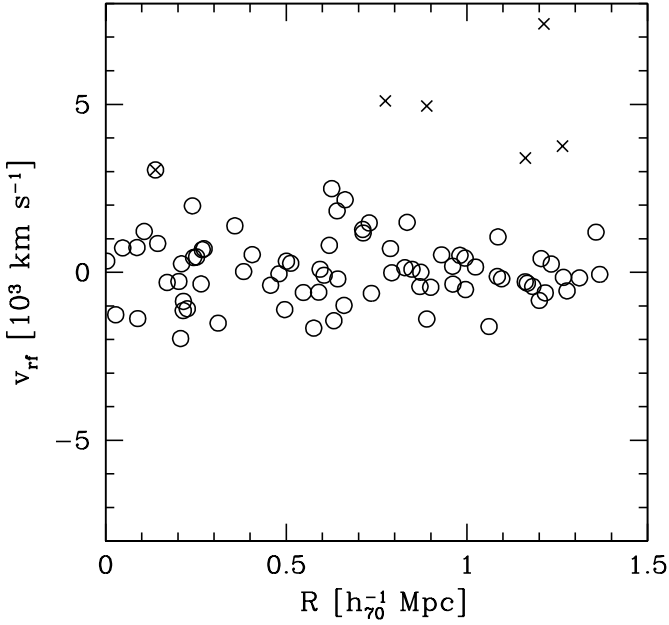}}
\caption
{The 83 galaxies assigned to the RX J105453.3+552102 density peak in
  Fig.~\ref{fighisto}.  Crosses indicate galaxies which were considered as interlopers on the basis of the gapper method. Circles indicate
  the 78 fiducial group members. The galaxy with the double symbol
  (i.e. SDSS J105454.00+552129.2) was not rejected to be more conservative. This choice did not affect the results of our dynamical analysis.}
\label{figvd}
\end{figure}

All the galaxies assigned to the group peak were further analyzed in the
second step which uses the combination of position and velocity
information: the ``shifting gapper'' method by Fadda et
al. (\cite{fad96}).  This procedure rejects galaxies that are too far
in velocity from the main body of galaxies within a fixed bin that
shifts along the distance from the group center.  The procedure is
iterated until the number of group members converges to a stable
value.  For the center of RX J105453.3+552102 we adopted the position of the BGG.  Fadda et
al. (\cite{fad96}) used a gap of $1000$ \ks in the cluster
rest--frame and a bin of either 0.6 \hh, or large enough to include 15
galaxies, these parameters well working in their cluster sample.  For
RX J105453.3+552102 this choice of the parameters rejects six galaxies (see
Fig.~\ref{figvd}). Out of these, the closest to the group center
(ID~SDSS J105454.00+552129.2) is very close to the main body of galaxies and would be
not rejected in the case of a small - comparable to typical $cz$-error - change
of the gap parameter (i.e. a gap of $1080$ \ks instead of $1000$
\kss). Moreover, the central velocity dispersion in a galaxy system is
often found to be larger than in outer regions (den Hartog \& Katgert \cite{den96}; Rines et al. \cite{rin03}; Aguerri et al. \cite{aguerri07}).  For the above reasons, we preferred to
be more conservative and did not reject the galaxy ID~SDSS J105454.00+552129.2.  This
leads to a sample of 78 fiducial members.  We verified that the
exclusion/inclusion of the above galaxy does not change the results of
our dynamical analysis.

\section{Photometric and spectroscopic luminosity functions of the group}

\begin{figure}
\centering
\resizebox{\hsize}{!}{\includegraphics{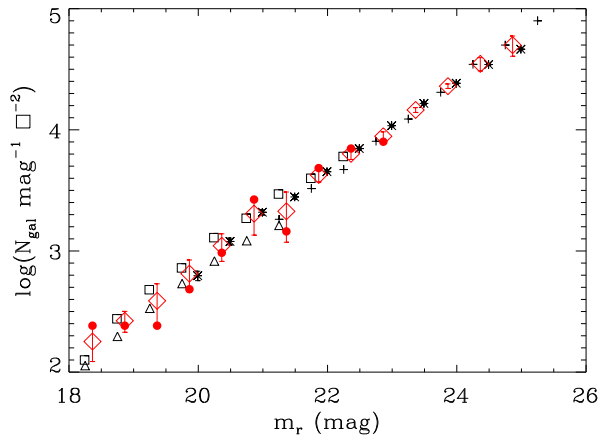}}
\caption
{Background galaxy number counts from: our control field (full points), Capak et al. (2004) (asterisks), Yasuda et al. (2001) (triangles), Huang et al. (2001) (squares), and Metcalfe et al. (2001) (crosses). The diamonds represent the mean background used for the photometric luminosity function.}
\label{background}
\end{figure}

We have computed the photometric and spectroscopic luminosity functions of RX J105453.3+552102. The photometric luminosity function (LF$_{\rm phot}$) was obtained as the statistical difference between the galaxy counts in the group and control field samples. The selection of the control field is critical since it will influence in the measurements of the parameters of the LF$_{\rm phot}$. In order to have a robust determination of the parameters of the LF$_{\rm phot}$, several control field samples were used. One of these control fields was observed by us the same night of the observations under the same instrumental and atmospheric conditions as the scientific images. However, this control sample is not so deep as the scientific image. Indeed, it limiting magnitude $m_{r}\sim 23$ mag, about 1 mag brighter than the limiting magnitude of the scientific image. Therefore we have also used deeper backgrounds than our control field from the literature: Capak et al. (\cite{capak04}), Yasuda et al. (\cite{yasuda01}), Huang et al. (\cite{huang01}), and Metcalfe et al. (\cite{metcalfe01}). All these control fields allowed us to obtain a statistical subtraction of the background galaxy counts down to our limiting magnitude. Figure~\ref{background} shows all the different backgrounds used in this work as well as the mean background which was used for the computation of the LF$_{\rm phot}$.

\begin{figure}
\centering
\resizebox{\hsize}{!}{\includegraphics{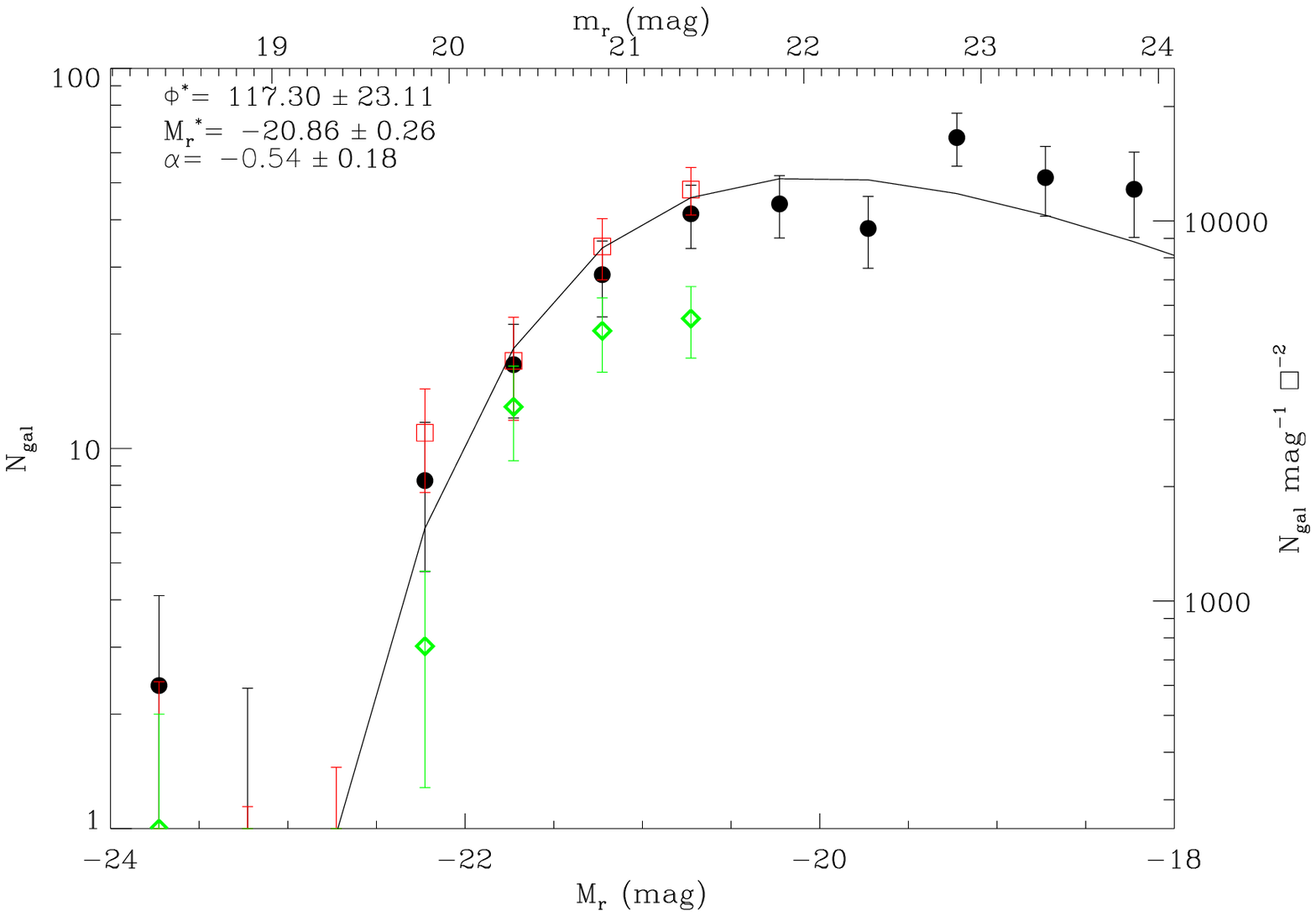}}
\resizebox{\hsize}{!}{\includegraphics{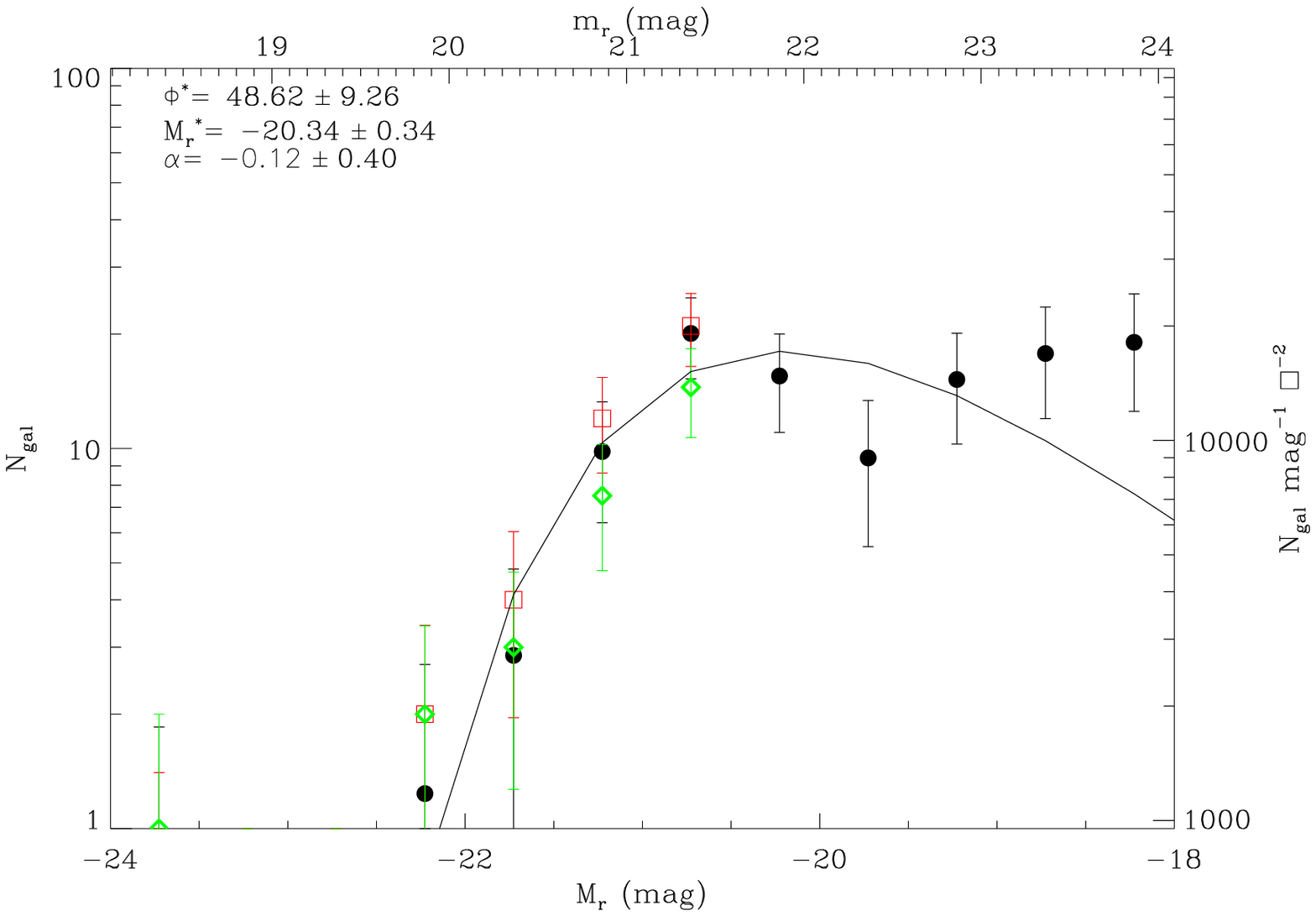}}
\caption
{Luminosity function of the galaxies of RX J105453.3+552102 within 1 Mpc (top panel) or 0.5 Mpc (bottom panel) radius, respectively. In all panels, the black points represent $\rm LF_{\rm phot}$ calculated by the statistical subtraction of galaxy counts obtained from the literature. The red squares represent $\rm LF_{\rm phot}$ in which the background counts were obtained using the photometric redshift from SDSS. The green diamonds represent $\rm LF_{\rm spec}$.}
\label{lf}
\end{figure}

Figure~\ref{lf} shows the photometric luminosity function of RX J105453.3+552102 obtained as the statistical difference between the galaxy counts of the group and the background galaxy counts (full black points). Notice that the photometric information for the group galaxies goes down to $\sim5$ magnitudes fainter than the bright cut-off of the LF.  This is one of the deepest photometric LFs of galaxies in a cluster at z$\sim 0.5$ (see also Rudnick et al. \cite{rudnick09}).

The galaxy LF$_{\rm phot}$ shown in Fig.~\ref{lf} was determined by using all galaxies within 1 and 0.5 Mpc radius from the BGG, respectively. Notice  that in the brightest magnitude bin of the LF computed with galaxies within 1 Mpc radius there is more than one galaxy after the statistical subtraction of the background galaxies. Two of these bright galaxies are located at a distance larger than 500 kpc from the center of the group. This do not invalid the selection of this group as FG. Indeed, according to the SDSS photometric redshifts these two bright galaxies are foreground objects. The galaxy LF$_{\rm phot}$ was fitted with a Schechter function (Schechter \cite{schechter76}):

\begin{equation}
\phi(m_{r})=\phi^{*} \times [10^{0.4(m^{*}-m_{r})}]^{\alpha+1}e^{-10^{0.4(m^{*}-m_{r})}},
\end{equation}

\noindent where $\alpha$ is the slope of the faint end of the LF, $m^{*}$ is the characteristic magnitude and $\phi^{*}$ is a normalization factor. The fit of the Schechter function to the data was done by minimizing the $\chi^{2}$ by taking errors into account and assigning a statistical weight to each of the points. The best-fit parameters are written in Fig.~\ref{lf}.

In order to check the strength of our LF$_{\rm phot}$ we have computed another LF using a different technique. In this case, members were selected using the photometric redshifts provided by SDSS. We have downloaded a catalogue with all the galaxies detected by SDSS-DR7 within a radius of 1 Mpc around the BGG of RX J105453.3+552102. We considered as foreground and background galaxies those with $z_{\rm phot}<0.37$ or $z_{\rm phot}>0.57$, respectively. The limits in the $z_{\rm phot}$ were chosen taking into account that the typical errors of SDSS $z_{\rm phot}$ are $\sim 0.1$. This field galaxy sample was used for computing a new photometric LF (red squares in Fig.~\ref{lf}; hereafter LF$_{\rm SDSS}$). Notice the agreement between LF$_{\rm SDSS}$ and LF$_{\rm phot}$.  

The spectroscopic luminosity function (LF$_{\rm spec}$) was determined using the velocities measured from our MOS observations. This LF was computed taking into account the group members and the selection function (i.e., the number of galaxies with redshift information over the number of possible spectroscopic targets per magnitude bin). Figure ~\ref{lf} shows LF$_{\rm spec}$ (green diamonds). Notice the good agreement within the errors between LF$_{\rm phot}$, LF$_{\rm SDSS}$, and LF$_{\rm spec}$ down to $M_{\rm r}\sim -21$, the limiting magnitude of our LF$_{\rm spec}$. 

We have integrated the fitted Schechter function in the $r$-band in order to obtain the total luminosity of the group. We have added the luminosity of the BGG, since this galaxy was not considered in the fit of the LF$_{\rm phot}$.Thus, the total luminosity of the group, within 1 Mpc radius, turned to be $L_{r}=2.0 \times 10^{12}$ L$_{\odot}$. The BGG accounts for $14\%$ of the total group luminosity. This percentage increases up to $32\%$ when only galaxies within 0.5 Mpc from the group center were considered.

Figure~\ref{lf} also shows a prominent dip in the LF$_{\rm phot}$ at $M_{r}\sim-19.5$. The dip is more clear in the LF$_{\rm phot}$ computed with the galaxies within 0.5 Mpc radius. This indicates that there is a lack of galaxies with $M_{r}\sim -19.5$ in the innermost regions of the group.

The photometric LFs are very sensitive to the background subtraction. In particular, the mean background used in the present work could affect to the dip detected in the LF$_{\rm phot}$. In order to check the dependence of this dip with the galaxy background used we have obtained the photometric LF of the group using only our own background. This background was taken in similar conditions as the observations and is deep enough in order to see the presence of the dip. This new LF$_{\rm phot}$ also showed the dip at $M_{r}\sim-19.5$. This indicates that the dip does not depend on the adopted galaxy background.

\section{Photometric properties of the brightest group galaxy}

\begin{figure*}
\centering
\resizebox{\hsize}{!}{\includegraphics{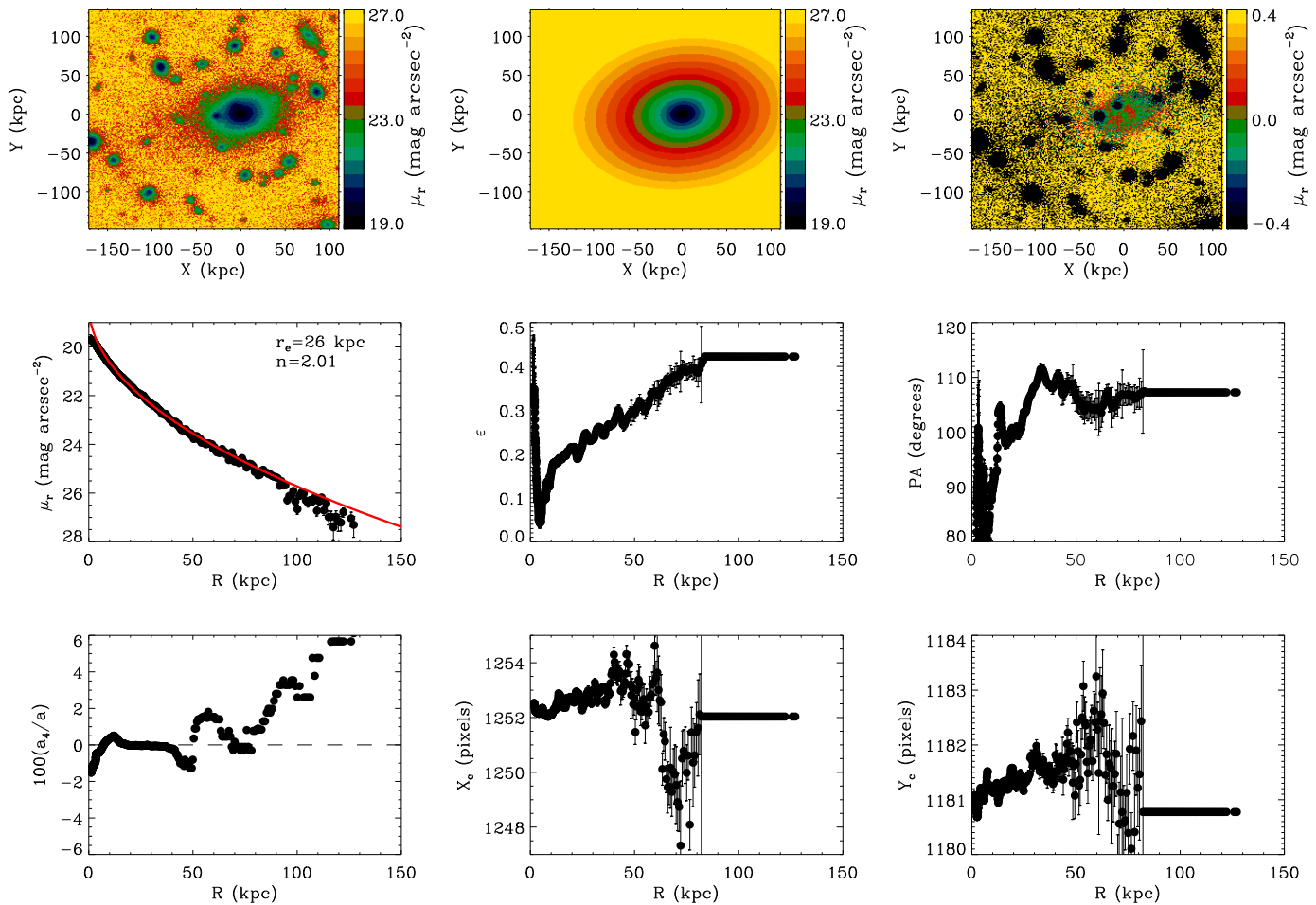}}
\caption
{Photometric properties of the BGG of RX J105453.3+552102. Top panels: $r$-band image of the BGG (left), 2D S\`ersic fitted model (center), and residuals of the fit (right). Middle panels: isophotal $r$-band dimming corrected surface brightness and unconvolved S\`ersic fitted model (left), ellipticity (center) and position angle (right) radial profiles of the BGG. Bottom panels: isophotal $a_{4}$ parameter (left), X coordinate (center), and Y coordinate (right) of the center of isophotes as a function of the galactrocentric distance.}
\label{photbgg}
\end{figure*}

We have fitted ellipses to the isophotes of the brightest group galaxy of RX J105453.3+552102. The fits were done using the {\small ELLIPSE} task from the IRAF package which uses the iterative algorithm described by Jedrzejewski (\cite{jedr87}). Figure ~\ref{photbgg} shows the $r$-band surface brightness (dimming corrected), ellipticity, and position angle radial profiles of the fitted ellipses. Our surface photometry extends out to $\sim 130$ kpc from the galaxy center. At this distance, the $r$-band surface brightness of the BGG is $\mu_{r}\sim27.0$ mag arcsec$^{-2}$. The ellipse fitting algorithm converged for galactrocentric radius smaller than $\sim 80$ kpc. The surface brightness of the isophotes with larger radius was obtained by imposing that the ellipticity and position angle were equal to those of the isophote with 80 kpc. These values were 0.42 and 106.8$^{\circ}$ for the ellipticity and position angle, respectively. Figure ~\ref{photbgg} also shows that the isophotal ellipticity profile of the BGG increases with radius and there is a twist of $\sim 10^{\circ}$ between the inner isophotes ($R>$10 kpc) and the outer ones ($R>40$ kpc). The difference between the inner and outer regions of the galaxy can be also seen in the change of the coordinates of the isophote center at radii larger than $\sim 60$ kpc (see Fig. ~\ref{photbgg}). This change in the isophotal center could be due to some tidal distortion in the outermost regions of the BGG.

Figure ~\ref{photbgg} also shows the radial profile of the Fourier coefficient $a_{4}$ of the isophotes of the BGG. This coefficient is related with the isophotes shape. Negative values of $a_{4}$/$a$ indicates boxy isophotes. In contrast, $a_{4}/a>0$ is related with discy ones. We can see that the $a_{4}/a$ radial profile of the BGG of RX J105453.3+552102 is close to zero in the inner $\approx 80$ kpc, and takes positive values in the outermost regions of the galaxy (R$>$80 kpc).  This parameter also indicates different photometric properties of the internal and external regions of the galaxy

We have fitted a two-dimensional S\`ersic model to the surface brightness of the BGG. The fit was done using the automatic fitting routine GASP2D (M\'endez-Abreu et al. \cite{mendez08}). Figure ~\ref{photbgg} shows the fitted model and the residuals of the fit. We have also overploted in Fig. ~\ref{photbgg} the one-dimensional fitted S\`ersic surface brightness profile to the isophotal surface brightness profile. Notice that the agreement between observations and model is remarkable. The surface brightness profile of the BGG is well fitted with a single S\`ersic profile. This fit reports a total magnitude of $m_{r}=17.49$ for the BGG. There is no extra light over the S\`ersic profile in the outermost regions of the galaxy. This indicates that this galaxy is not a cD galaxy (see Graham et al. \cite{graham96}; Nelson et al. \cite{nelson02}; Gonzalez et al. \cite{gonzalez05}; Patel et al. \cite{patel06}; Seigar et al. \cite{seigar07}; Vikram et al. \cite{vikram10}). It is also interesting to note that the fitted S\`ersic profile has $n\sim2$. According to the magnitude-$n$ relation follow by spheroidal galaxies we should expect a larger value of $n$ (see e.g., Aguerri et al. \cite{aguerri04}, and references therein). Recently, it has been found that the S\`ersic shape parameter of massive early-type galaxies is smaller for galaxies at higher redshift (van Dokkum et al. \cite{vandokkum10}; Vikram et al. \cite{vikram10}). Nevertheless, at $z\sim0.5$ there are no galaxies in the sample by van Dokkum et al. (2010) with $n\sim2$. Only galaxies at $z\sim2$ and extremely compact show values of $n\sim2$ (see fig. 7 van Dokkum et al. \cite{vandokkum10}). These structural differences of the BGG with respect to other bright early-type galaxies could indicates a different formation. In particular, the low value of $n$ unveil the kind of mergers that have formed the galaxy. In order to get a similar light profile it is required a very gas rich merger for the progenitors ($\sim 80\%$ gas rich; see Hopkins et al. \cite{hopkins08}).

\section{Internal dynamics}
\label{anal}

\subsection{Global dynamical properties}
\label{glob}

\begin{figure}
\centering
\resizebox{\hsize}{!}{\includegraphics{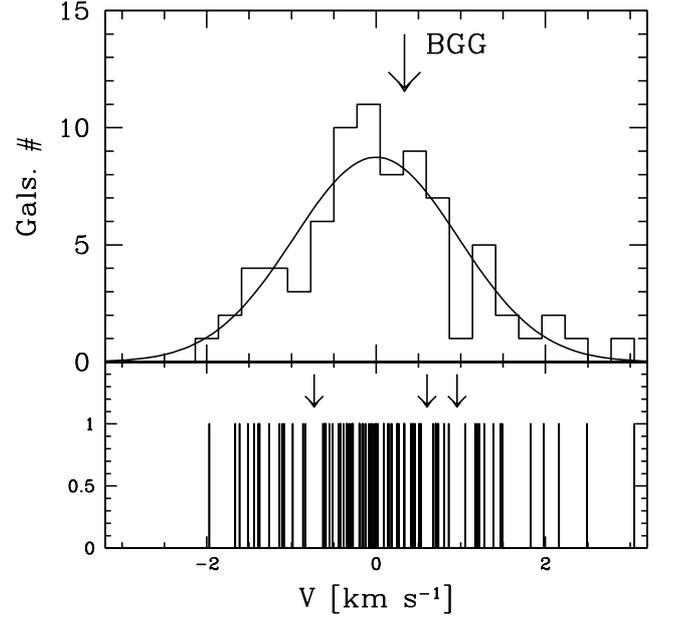}}
\caption
{{\em Upper panel}: Rest--frame velocity histogram for
  the 78
  group members. The arrow indicates the
  velocity of the BGG.
{\em Lower panel:}
Stripe density plot where the arrows indicate the positions of the
significant gaps.
}
\label{figstrip}
\end{figure}

Figure~\ref{figstrip} shows the velocity distribution of the 78 member galaxies. By applying the bi-weight estimator to the 78 member galaxies (ROSTAT
package; Beers et al. \cite{bee90}), we computed a mean redshift
of $\left<z\right>=0.4661\pm$ 0.0004, i.e.
$\left<\rm{v}\right>=(139731\pm$110) \kss.  We estimateed the line--of--sight (LOS) velocity dispersion, $\sigma_{v}$, by using the bi-weight estimator
and applying the cosmological correction and the standard correction
for velocity errors (Danese et al. \cite{dan80}).  We have obtained
$\sigma_{v}=969_{-85}^{+107}$ \kss, where errors were estimated
through a bootstrap technique. This velocity dispersion does not significantly change when we exclude the six blue galaxies (m$_{g}$-m$_{i} < 1.5$). In this case, $\sigma_{v}=993_{-87}^{+119}$ \kss.

\begin{figure}
\centering
\resizebox{\hsize}{!}{\includegraphics{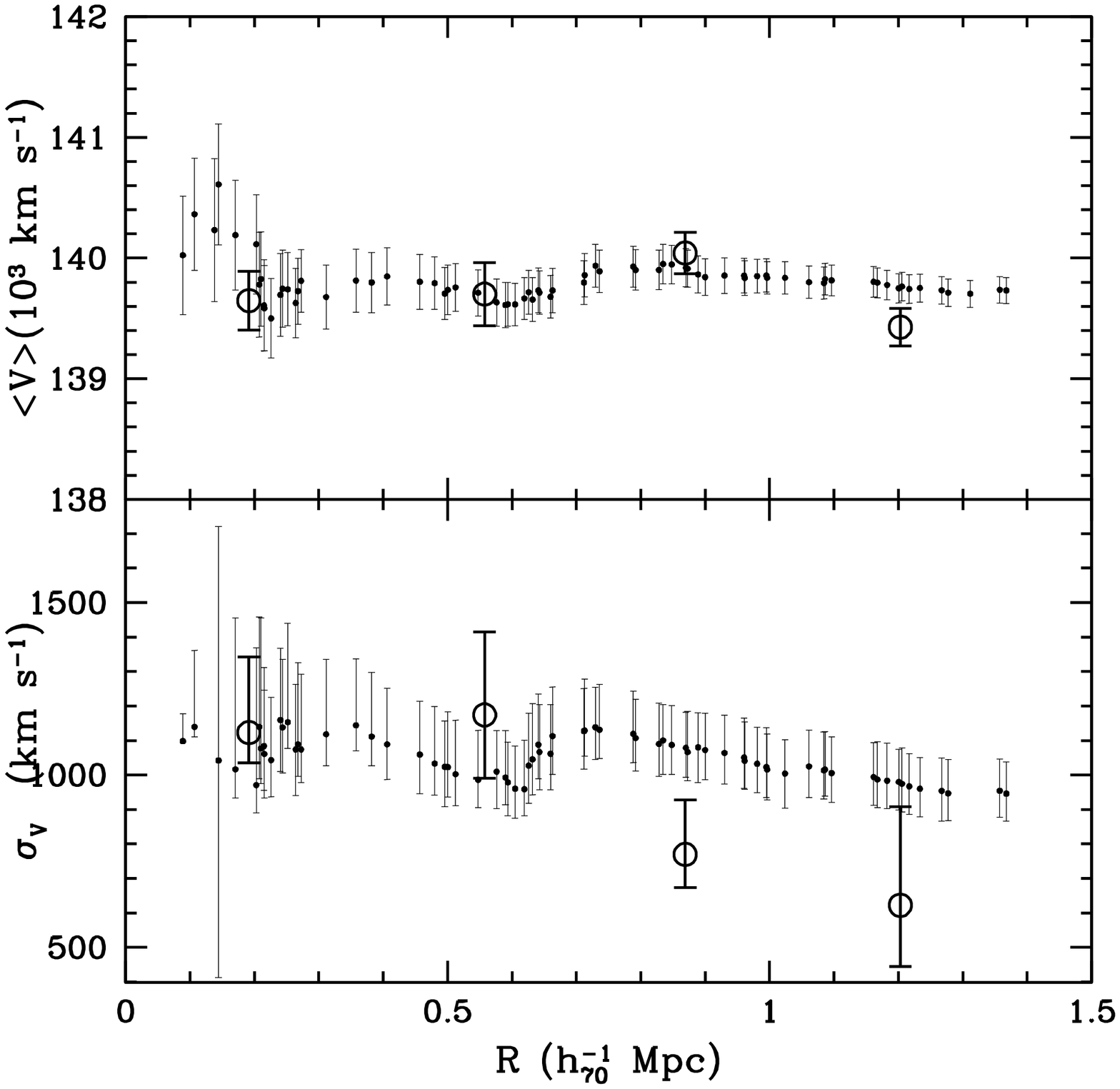}}
\caption
{ Differential (big circles) and integral (small points) profiles of
  mean velocity ({\em upper panel}) and LOS velocity dispersion ({\em
    lower panel}).  The differential radial profiles were obtained by averaging velocities and velocity dispersions in four 0.35 \h wide annuli. The integral radial profiles were obtained by averaging velocities and velocity dispersions within each (projected) radius. The first values are computed on the five galaxies closest to the center. The errors are the $68\%$ c.l.}
\label{figprof}
\end{figure}

To evaluate the robustness of the $\sigma_{v}$ estimate we analyzed
the velocity dispersion profile (Fig.~\ref{figprof}).  The integral
pigrofile is almost flat. This indicates that a robust value of
$\sigma_{v}$ is already reached in the internal regions (e.g.,
Fadda et al. \cite{fad96}; Girardi et al. \cite{gir96}).  

In the framework of usual assumptions: i) system sphericity; ii)
dynamical equilibrium; iii) the galaxy distribution traces the
mass distribution, one can compute virial global quantities. Following
the prescriptions of Girardi \& Mezzetti (\cite{gir01}), we assumed for
the radius of the quasi--virialized region $R_{\rm
  vir}=0.17 \times (\sigma_{v}/{\rm km~s^{-1}}) / H(z) \times {\rm Mpc}= 1.83$ \h (see their Eq.~1
with the scaling with $H(z)$; see also the Eq.~ 8 of Carlberg et
al. \cite{car97} for $R_{200}$).  We have redshifts for galaxies
out to a radius of $R_{\rm out}\sim 1.37 $ \h sampling the region within
$\sim0.76\times R_{\rm vir}$.

One can compute the virial mass (Limber \& Mathews \cite{lim60}; see
also, e.g., Girardi et al. \cite{gir98}) using the data for the $N_{\rm g}$
observed galaxies:

\begin{equation}
M=3\pi/2 \cdot \sigma_{v}^2 R_{\rm PV}/G-\rm{SPT} ,
\end{equation}

\noindent where
SPT is the surface pressure term correction (The \& White
\cite{the86}), and
$R_{\rm PV}$, equal to two times the (projected) mean harmonic radius, is: 

\begin{equation}
R_{\rm PV}=\frac{N_{\rm g}(N_{\rm g-1})}
{\Sigma_{\rm j< \rm i}\Sigma_{\rm i}R_{\rm ij}^{-1}},
\end{equation}

\noindent 
where $R_{\rm ij}$ is the projected distance between two galaxies.

The estimate of $\sigma_{v}$ is a robust estimate in RX J105453.3+552102 (see
Fig.~\ref{figprof}) and thus we considered our global value.  The value
of $R_{\rm PV}$ depends on the size of the sampled region and possibly
on the quality of the spatial sampling, e.g. whether the galaxy sample
suffer of radial-dependent incompleteness.  In particular, ignoring
the problem of radial-dependent incompleteness can have quite a
catastrophic effect on the cluster mass estimate (see Sect.~4.4 of
Biviano et al. \cite{biv06}).  We verified that our sample is not biased
comparing the distribution of cluster-centric distances of galaxies
having redshift vs. those of all SDSS galaxies having $0.37<z_{\rm
  phot}<0.57$ (no difference according to the Kolmogorov-Smirnov test,
hereafter 1DKS--test; see e.g., Press et al. \cite{pre92}).  For our
sampled region, i.e.  within $R_{\rm out}$ we obtain
$R_{\rm PV}=(1.21$$\pm$0.09) \hh, where the error was obtained via a
jackknife procedure (see e.g., Press et al. 1992).  The value of SPT correction strongly depends on
the amount of the radial component of the velocity dispersion at the
radius of the considered region and could be obtained by analyzing the
velocity--dispersion profile, although this procedure would require
several hundreds of galaxies.  Combining data of many clusters it results that velocities are isotropic and that the SPT correction at
$\sim R_{\rm 200}$ is $\rm{SPT}/M_{\rm v}\sim20\%$ (e.g., Carlberg et
al. \cite{car97}; Girardi et al. \cite{gir98}). Applying the same
correction to our mass estimate we obtained $M(<R_{\rm out})=(1.0\pm 0.2)$ \mquii.

To obtain the mass within the whole virialized region, which is larger
than that sampled by observations we used an alternative estimate
of $R_{\rm PV}$ on the basis of the knowledge of the galaxy-number
distribution.  Following Girardi et al. (\cite{gir98}; see also their
approximation given by Eq.~13 when $A=R_{\rm vir}$) we assumed a
King--like profile with parameters typical of galaxy
clusters: a core radius $R_{\rm c}=1/20\times R_{\rm vir}$
and a slope--parameter $\beta_{\rm fit}=0.8$, i.e. the volume
galaxy-number density at large radii goes as $r^{-3
  \beta_{\rm fit}}=r^{-2.4}$. Notice that the same values of $R_{\rm
  c}$ and $\beta_{\rm fit}$ were also found by directly fitting the data of
RX J105453.3+552102.  We
obtained $R_{\rm PV}(<R_{\rm vir})=1.36$ \hh, where a $25\%$
error was expected (Girardi et al. \cite{gir98}).  Assuming
the $20\%$ SPT correction we computed $M(<R_{\rm vir})=1.1_{-0.3}^{+0.4}$ \mquii.

Adopting a completely different approach based on numerical N-body
simulations, we used the Eq. 2 of Biviano et al. (\cite{biv06}),
opportunely rescaled for cluster redshift and our cosmology 

\begin{equation}
M_{\rm v}(<R_{\rm v})=11.13 (\sigma_{v}/10^3{\rm km\ s^{-1}})^3 \times H_0/H(z) \mquaaa,  
\end{equation}

\noindent 
to compute a mass estimate $M_{\rm v}(<R_{\rm v}=1.62 \hhh)=7.9$
\mquaa, where $R_{\rm v}$ is another estimate for $R_{200}$. Notice that, although this and the above empirical procedure
differ in fixing the value of $R_{\rm 200}$, the two mass
estimates only differ for $\sim 10\%$ when rescaled to the same
radius.

\subsection{Velocity distribution}
\label{velo}

We have analyzed the velocity distribution to look for possible deviations
from Gaussianity that can be interpreted as the signature of a complex
dynamics. For the following tests the null hypothesis is that the
velocity distribution is a single Gaussian.

We estimated three shape estimators, i.e. the kurtosis (KURT), the skewness (SKEW),
and the scaled tail index (STI) using the ROSTAT package and we computed the
significance levels according to Beers et al. (\cite{bee91}; see their
Table~2).  The scale tail index (STI=1.152) and the normalized
kurtosis (KURT=0.370) turned to be consistent with a Gaussian. In contrast, the skewness
(SKEW=0.454) shows evidence of departure from a Gaussian (with
a c.l. in the $95-99\%$ range).  However, as noticed by Beers et
al.~(\cite{bee91}), the STI index is a conservative diagnostic for
studying galaxy clusters, while the coefficients of skewness and
kurtosis are biased toward falsely significant values. In our case, the
rejection of only one galaxy (i.e., SDSS J105454.00+552129.2) which is at the border line
of our member selection (see \S~\ref{memb}), leads to SKEW=0.242 which is fully
consistent with a Gaussian velocity distribution. 

It is possible to interpret the shape of the velocity distribution as a rough indicator of the orbital anisotropy of the group members. Thus, the halo of a dynamically hot system with isotropic orbits and a constant circular velocity curve (i.e., with a massive dark matter halo) has a Gaussian LOS velocity distribution and KURT=0 (see Gerhard 1993). Instead, a system containing objects on radial orbits tend to produce a centrally peaked LOS velocity distribution with long tails, resulting in a positive value of kurtosis. As discussed earlier, the degree of fossilness might reside in the presence of radial orbits in the outer group regions. For this reason we have attempted to look for signatures of non-Gaussianity in the group outskirts by dividing the spectroscopic sample in three radial bins (with 26 members each), and recomputed the kurtosis parameter for each bin. Due to the small number statistics, we have used a bootstrap approach to carefully take into account the effect of the binning and the presence of outliers. The final result is that the first two radial bins ($R\sim0.5', 1.5'$) the kurtosis turned out to be still fully consistent with a Gaussian distribution, while in the third bin, which includes objects with $R> 2'$, we have found KURT$=1.0^{+0.9}_{-0.7}$ with errors accounting 10$\%$ and 90$\%$ quartiles over 500 bootstrap experiments\footnote{Each bootstrap experiment consists in randomly picking a number of velocity values in the bin and substituting with a random velocity taken from a Gaussian distribution with mean and standard deviation of the original velocity distribution in the bin and then recompute the kurtosis. This implicitly forces the post-bootstrap kurtosis toward a value which is closer to a Gaussian one, which makes our conclusion of a non-Gaussianity of the distribution even stronger}. This results suggest that there is a statistically significant departure from Gaussianity of the group member velocities outside $R=2'$, which we interpret as a hint of radial orbits at these radii. However a full assessment on the actual amount of anisotropy of the system will require a more detailed dynamical analysis, which is beyond the purpose of this paper.

We also investigated the presence of gaps in the velocity distribution.
We followed the weighted gap analysis presented by Beers et
al. (\cite{bee91}; \cite{bee92}; ROSTAT package).  We looked for
normalized gaps larger than 2.25 since in random draws of a Gaussian
distribution they arise at most in about $3\%$ of the cases,
independent of the sample size (Wainer \& Schacht~\cite{wai78}).  We
detected three significant gaps (at the $97\%$ c.l.) which divide the
system in four subgroups of 14, 44, 7 and 13 galaxies from low to high
velocities (hereafter GV1, GV2, GV3 and GV4).  The BGG was
assigned to the GV2 peak (see Fig.~\ref{figstrip}).  The above
probabilities are ``per gap'' probabilities.  The probability of
finding gaps of this large somewhere in the distribution is $69\%$,
i.e. the ``cumulative'' probability was not statistically
significant.

Following Ashman et al. (\cite{ash94}) we also applied the Kaye's
mixture model (KMM) algorithm which fits a user--specified number of
Gaussian distributions to a data-set and assesses the improvement of
that fit over a single Gaussian.  We adopted the results of the gap
analysis to determine the first guess for the group partition.  We
found no indication that a four--groups partition  is a
significantly better descriptor of the velocity distribution with
respect to a single Gaussian.

Finally, we checked the presence of a significant velocity offset of the
BGG with respect to the central location in the velocity space of the
remaining system galaxies.  We followed the approach of the bootstrap
test by Gebhardt \& Beers (\cite{geb91}) who criticized previous
methods suggesting a more rigorous approach. By adopting the opportune
biweight estimators used above and the rest-frame correction we obtained
that $Z=(V_{\rm BGG}- \left<\rm{v}\right>)/\sigma_{v}=0.344$, being $V_{\rm BGG}$ the radial velocity of the BGG.  The
associated bootstrap intervals of $Z$ were [0.130,0.569],[0.078,0.605],
[0.000,0.682] at the $90\%$, $95\%$, $99\%$ c.l., respectively.
Thus, the zero offset can be excluded at $>95\%$ c.l. The
same result was obtained taking into account a $90\%$ error on $V_{\rm
  BGG}$ due to its redshift uncertainty ([0.072,0.614],[0.020,0.654],
[0.000,0.738] at the $90\%$, $95\%$, $99\%$ c.l.).

\subsection{2D analysis}
\label{2D}

While the tests for Gaussianity are useful for detecting mergers
occurring along our LOS, where the velocity distribution is
significantly perturbed, they could be ineffective in the case of a
merger occurring near the plane of the sky (Pinkney et
al. \cite{pin96}).  In these cases, galaxy density maps are useful
tools for searching for evidence of substructure.

\begin{figure}
\centering
\includegraphics[width=8cm]{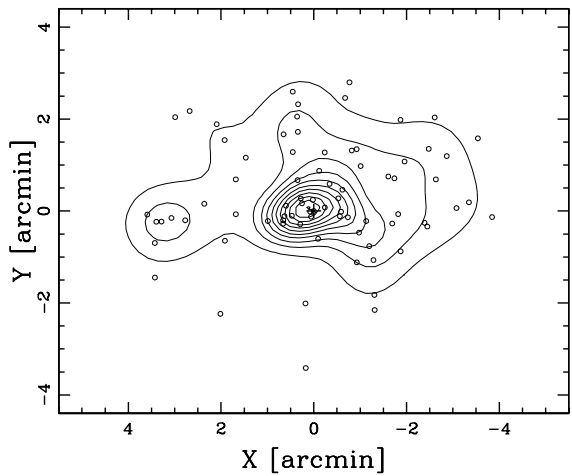}
\caption
{Spatial distribution on the sky and relative isodensity contour map
  of RX J105453.3+552102 members, obtained with the DEDICA method.  The plot is
  centered on the BGG, indicated by a cross. The eastern minor peak
is not significant.}
\label{figk2z}
\end{figure}

When applying the DEDICA method to the 2D distribution of the 78
galaxy members we found a main, very significant peak and a minor
eastern peak (see Fig.~\ref{figk2z}). The main peak is well centered
on the BGG.  

The secondary peak is much less significant than the main
peak. Although the statistical significance given by DEDICA procedure is
very useful to rank the importance of the subclusters, its physical
meaning should be discussed.  Ramella et al. (\cite{ram07}) tested the
2D DEDICA procedure on Monte-Carlo simulations reproducing galaxy
clusters. They showed that the physical significance associated to the
subclusters is based on the statistical significance of the subcluster
(recovered from the $\chi ^2$ value) and the $r_{cs}=N_{\rm C}/N_{\rm
  S}$ parameter, where $N_{\rm S}$ is the number of members of the
substructure and $N_{\rm C}+N_{\rm S}$ is the total number of cluster
members.  For the secondary peak of RX J105453.3+552102 we computed $\chi ^2=5.4$
and $r_{cs}=6.1$.  Figure~2 of Ramella et al. shows how a
subcluster with the above $\chi ^2$ has a large probability ($>20\%$)
to be due to the simulated noise fluctuations, while a real subcluster
with $r_{cs}\sim 6$ is expected to have a $\chi ^2$ in the range
$20-30$. Thus we concluded that the secondary peak in our 
Fig.~\ref{figk2z} is likely to be a false positive event.

\begin{figure}
\includegraphics[width=8cm]{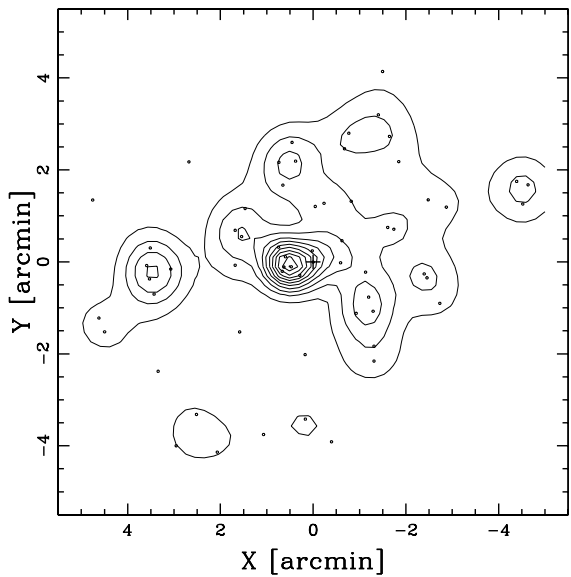}
\caption
{Spatial distribution on the sky and relative isodensity contour map
  of the ``likely'' red cluster members extracted from the 
 SDSS photometric catalogue.} 
\label{figk2ccrvir}
\end{figure}

To overcome possible biases connected with the spatial coverage
  of the spectroscopic data and to make a homogeneous analysis in the
  whole virial region, we resort to the photometric catalogue extracted
  from the SDSS considering all galaxies with $0.37<z_{phot}<0.57$
  within $R_{\rm vir}$. To further limit possible contamination by
  field galaxies we analyzed only "red" galaxies performing an
  additional selection on the base of the ($r$--$i$) -- ($i$--$z$)
  colour--colour plane (see e.g. Goto et al. \cite{got02}; Boschin et
  al. \cite{bos08}).  The median values of SDSS $r$--$i$ and $i$--$z$
  colours of the spectroscopically cluster members were 0.79 and 0.40
  mag, respectively.  Following Goto et al. (\cite{got02}, see their fig.~12), we
  selected a rectangular window around these median values, here
  asymmetric to limit the field contamination, ($0.7 \leq r$--$i\leq
  1.0$ and $0.2\leq i$--$z\leq 0.6$).  By applying this selection
  criterium to the photometric SDSS catalogue, we obtained a sample of 61
  ``likely'' cluster members, 36 and 3 of which are spectroscopically
  members and non members, respectively. Figure~\ref{figk2ccrvir}
  shows the result of the application of DEDICA to the likely
  members. As in the case of the spectroscopic members, we found one,
  very pronunced density peak.

\subsection{3D structure}
\label{3d}

The existence of correlation between position and velocity of
cluster galaxies is a footprint of real substructure. i.e., of
  galaxy subsystems which reside within the galaxy cluster. Here, we
used three different approaches to analyze the structure of RX
J105453.3+552102 combining velocity and position information.

To check whether the weighted gaps detected in \S~\ref{velo} have a
physical meaning we compared two by two the spatial galaxy
distributions of GV1, GV2, GV3 and GV4 by using the 2D
Kolmogorov--Smirnov test (hereafter 2DKS--tests; Fasano \&
Franceschini \cite{fas87}, as implemented by Press et
al.~\cite{pre92}). We found no difference, and therefore no
support to the existence of these four sub-clumps in RX J105453.3+552102.

The cluster velocity field may be influenced by the presence of
internal substructures.  Following Girardi et al. (\cite{gir96}; see
also den Hartog \& Katgert \cite{den96}) we analyzed the presence of a
velocity gradient performing a multiple linear regression fit to the
observed velocities with respect to the galaxy positions in the plane
of the sky and assessed its significance performing 1000 Monte Carlo
simulations.  We found no significant velocity gradient.

We combined galaxy velocity and position information to compute the
$\Delta$--statistics devised by Dressler \& Schectman (\cite{dre88};
see, e.g., Boschin et al. \cite{bos09}, for a recent application). This
test is sensitive to spatially compact subsystems that have either an
average velocity that differs from the system mean, or a velocity
dispersion that differs from the global one, or both.  We found no
significant evidence of substructure.

\section{Discussion and conclusions}
\label{disc}

\subsection{Cluster global properties}

Our analysis has shown that RX J105453.3+552102 is a quite massive galaxy
system.  We estimated a velocity dispersion $\sigma_{v}\sim 1000$
\ks and a virial mass $M(<R_{\rm vir}) \sim 1.1$ \mquii, comparable
to the values estimated for the Coma cluster (e.g., Colless \& Dunn \cite{colles96}; Girardi et al. \cite{gir98}).  To date no measurement of X-ray
temperature is available.  In the assumption of density--energy
equipartition between gas and galaxies, i.e. $\beta_{\rm
  spec}=1$ \footnote {$\beta_{\rm spec}=\sigma_{v}^2/(kT/\mu
  m_{\rm p})$ with $\mu=0.58$ the mean molecular weight and $m_{\rm
    p}$ the proton mass.}, we expect $T_{\rm X} \sim 6$ keV.

RX J105453.3+552102 is a very luminous cluster. We estimated $L_{r}=2.0$ \ldod for
the $r$-band luminosity projected within 1 \hh. We used the King-like
profile for the galaxy-number distribution already adopted in
\S~\ref{glob} to deproject and extrapolate the observed luminosity (without the BGG)
and then computed the total luminosity (BGG+other galaxies)
$L_{r}=2.4$ \ldod within a sphere of $1R_{\rm vir}$ radius.  For
comparison with nearby clusters we must consider that $L_{r}(z=0)\sim
1.7 L_{r}(z=0.47)$ for the luminosity of early-type galaxies
($k_{r}\sim -1$ and $E_{r}=0.86*z\simeq +0.4$ for the k- and
evolutionary corrections, respectively; Fukugita et al. \cite{fuk95};
Roche et al.  \cite{roc09}), while small or no global correction is
due for the luminosity of spiral galaxies (e.g., Poggianti et
al. \cite{pog97}).  Thus $L_{r}\lesssim 4.1$ \ldod for a
corresponding cluster at $z=0$.  This value lies in the high tail of
luminosity distribution of cluster galaxies (RASS-SDSS sample, see
fig.~2 of Popesso et al. \cite{pop07b}).  

RX J105453.3+552102 is also quite X-ray luminous.  From the counts listed in the
RASS Faint Source Catalogue we estimated a X-ray luminosity 
$L_{\rm X(0.1-2.4)keV}\sim 4.8\times 10^{44}$ \lumx erg s$^{-1}$, where we
used the conversion factor by B\"ohringer et al. (\cite{boe00}; Fig.~8a)
  and roughly corrections for the missing flux and k-correction
  $10\%$ and $\lesssim 20\%$, respectively (B\"ohringer et
  al. \cite {boe00}; Mullis et al. \cite{mul03}). Assuming
the above estimated  $T_{\rm X}$ we obtained $L_{\rm X, bol}\sim 11\times
10^{44}$ \lumx erg s$^{-1}$.

The properties of RX J105453.3+552102 are well consistent with those of other,
typical clusters.  The value of the mass--to--light--ratio (for the
corresponding cluster at $z=0$) $M/L_{r}\gtrsim 270$ \ml is well
comparable to that of nearby clusters from SDSS (see fig. 9 of Popesso et al. \cite{pop07b} for a cluster with $M_{200}\sim 1$ \mquii).  The position of RX J105453.3+552102  in the $L_{\rm X,
  bol}$-$\sigma_{v}$ plane and $L_{\rm X, bol}$-$M$ plane is well
consistent with that of other clusters (see fig.~5 of Girardi \&
Mezzetti \cite{gir01} and fig.~5 of Ortiz-Gil et al. \cite{ort04},
taking into account the different cosmologies; Popesso et
al. \cite{pop07a}).

We also considered the relation between X--ray and optical luminosities.
This relation is particularly unclear for fossil systems.  Khosroshahi
et al. (\cite{khosro07}, and references therein) claimed that, for a
given optical luminosity of the group, FGs are more X--ray luminous
than non-fossil groups. This is also predicted by numerical N--body
simulations (D'Onghia et al. \cite{donghia05}).  However, Voevodkin et
al. (\cite{voevodkin10}), analyzing someway more massive systems, found that
there is no difference between the FGs and the other systems analyzed with the
same technique.  Thus, to date it is not clear whether the
difference between FGs and other groups is a result of possible
systematic differences or is real for less massive systems (Voevodkin et
al. \cite{voevodkin10}).  As for
RX J105453.3+552102, for comparison with Voevodkin et al. (\cite{voevodkin10}), we considered
its X-ray luminosity $L_{\rm X(0.5-2.0)keV} \sim 2.9\times 10^{44}$
\lumx erg s$^{-1}$ and its optical luminosity $L_{r,500}=2.0$ \ldod
estimated within a sphere of radius $R_{500}=1.4$ \h (according to the
authors definition of $R_{500}$).  The position of RX J105453.3+552102 in the plane
($L_{\rm X(0.5-2.0)keV}$, $L_{r,500}$) is well consistent with that
of other clusters.

\subsection {Does RX J105453.3+552102 follow the definition 
of a ``fossil group''?}

RX J105453.3+552102  hosts a very luminous BGG as shown by plot of $L_{\rm BGG}$ vs
$M_{200}$ (Popesso et al. \cite{pop07b}, Fig.~14); indeed, its
$L_{r}\sim 3\times 10^{11} h_{70}^{-2} L_{\odot}$  is one of the highest values among massive
clusters with mass $\sim 1$ \mquii. The BGG light fraction, defined
as the ratio of BGG luminosity-to-total cluster galaxy light (BGG and
other galaxies), is $\sim 0.15$ inside 1 Mpc radius. This value is much smaller than that claimed for a
few fossil groups (e.g. $\sim 0.7$ for RXJ1340.6+4018; Jones et al.
\cite{jones00}). However, notice that the BGG light fraction strongly
depends on the mass of the system, decreasing with cluster
mass as shown by Lin \& Mohr (\cite{lin04}, see their fig.~4).  According with its mass, RX J105453.3+552102  lies at the superior boundary of the locus
occupied by clusters with mass $\sim 1$ \mquii.

The classical FG definition is based on the magnitude gap between
    the BGG and the second ranked galaxy within 500 kpc radius (Jones
    et al. 2003). This definition takes into account the group membership. Thus, a galaxy group is fossil when $\Delta m_{12}>2$.
    The fossil group classification is clearly strongly dependent
    on the magnitude estimation of the bright galaxy group.
    Unfortunately, this is not an easy
    task due to the BGG is often located in high density galaxy
    environments.  Thus, there is a large difference between SDSS model ($m_{r,model}=17.69$) and  Petrosian magnitudes ($m_{r,petro}=18.10$) for the BGG of RX J105453.3+552102.  From our photometry, the
    magnitude of the BGG calculated by SExtractor is $m_{r,SEx}=18.08$, and the
    magnitude obtained from its surface brightness fit is
    $m_{r,fit}=17.49$. Notice the agreement between $m_{r,petro}$ and $m_{r,SEx}$ and between $m_{r,model}$ and $m_{r,fit}$. Nevertheless, the model magnitudes are always brighter than aperture ones because they are computed integrating until infitive radius the fitted surface brightness profiles of the galaxies. The 0.2 mag difference between $m_{r,model}$ and $m_{r,fit}$ could be due to the best S\'ersic fitted model by SDSS has $n=1$, while our best fitted model has $n\sim 2$. 

\begin{figure}
\centering 
\resizebox{\hsize}{!}{\includegraphics{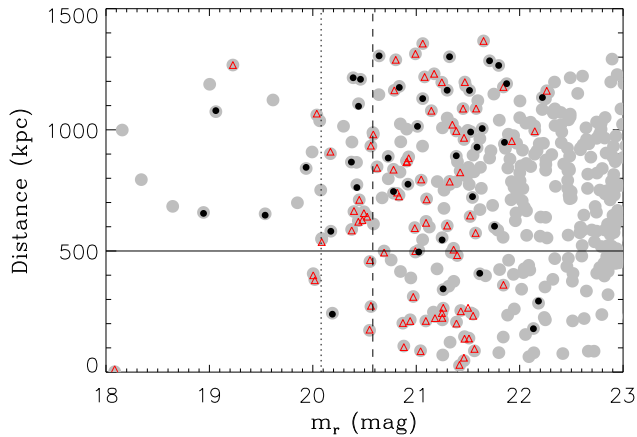}}
\caption{
Distance to the group center vs SExtractor r-band magnitude for all galaxies (grey circles). We have also overplotted cluster members (red triangles) and non-cluster members (black circles). The horizontal line shows 500 kpc distance from the galaxy group center (the distance used in the ``fossil group'' definition). The vertical dotted and dashed lines show $\Delta m_{12}=2.0$ and $\Delta m_{14}=2.5$ respectively, taking $m_{r,SEx}$ as the BGG magnitude.}
\label{colormag2}
\end{figure}

The differences between model and aperture magnitudes are crucial for comparing magnitudes of the same class. Thus, when considering our $m_{r,SEx}$ magnitudes of the cluster galaxies we obtained a magnitude gap between the BGG and the second rank galaxy within 500 kpc radius of $\Delta m_{12}=1.92\pm0.09$. This
    magnitude gap can be seen in Fig.~\ref{colormag2}. The value of 1.92 is very close to the
    classical FG definition given by Jones et al. (2003), but does not
    allow to classify RX J105453.3+552102 as a fossil group. Taking into account the errors there is a probability of $\sim 20\%$ to have $\Delta m_{12}>2$. We also computed from our photometry the model magnitudes of the second brightest galaxies of the group within 500 kpc radius. In this case $\Delta m_{12}=1.87\pm0.15$.

Recently, Dariush et al. (\cite{dariush10}) have proposed another
photometrical definition of FGs based on the magnitude gap between the
BGG and the fourth ranked galaxy ($\Delta m_{14}$). Analysing groups
and clusters of galaxies using the Millennium Simulation, they found
that early-formed galaxy association are better identified as those
showing $\Delta m_{14}>2.5$ mag. In our case, the RX J105453.3+552102
group has $\Delta m_{14}=2.47\pm0.09$ mag (using our $m_{r,SEx}$) and,
again, this group cannot be classified as fossil (see Fig.~\ref{colormag2}).  In this case, the probability that the system has $\Delta m_{14}>2.5$ is $\sim 35\%$. The same value of $\Delta m_{14}$ was obtained when $m_{r,fit}$ of the 4th brightest galaxy of the group within 500 kpc was considered.

As shown above, the classification of a system as fossil or not can be
quite sensible to the BGG magnitude estimation or to the presence of
bright interlopers in the cluster field.  More in general, the
classification scheme of a fossil group might be improved in several
ways, e.g., taking into account a radius scaling with $R_{\rm 200}$
and a magnitude band changing with redshift. However, the discussion
of this scheme is out of the aims of FOGO project which are rather to
check of how many groups in the catalog of Santos et
al. (\cite{santos07}) actually have ''fossil'' nature and to study their
properties. As for RX J105453.3+552102, its real nature, the likely
past dynamical history, and properties are discussed in the next
sections.

\subsection{The dynamical state of the cluster}

The presence of a large magnitude gap has been always taken as indication
of relaxed and early-formed galaxy systems. Nevertheless, in the
Millennium Simulation can be seen that most of the early-formed
systems do not show large magnitude gaps (see Dariush et
al. \cite{dariush10}). Thus, overcoming the empirical definitions of
``fossil group'', one should consider whether RX J105453.3+552102 is
or is not an old and undisturbed system that has underegone little infall
of $L^*$ galaxies since its initial collapse.

Recent major mergers with other galaxy systems would yield some
observable smoking-guns.  The first would be the presence of
substructure in RX J105453.3+552102.  We have used a battery of
different tests in 1D, 2D and 3D to take into account the geometry of
a possible cluster merger (Pinkney et al. \cite{pin96}). We found no
evidence of substructure.  The only possible hint is the peculiar
velocity of the BGG galaxy (significant at the $>95\%$ c.l.), which is
often connected to evidence of substructure (e.g. Bird \cite{bir94}).
However, the velocity of the BGG in the cluster rest frame is only
$\sim 300$ \ks and the relative peculiar velocity with respect to the
cluster velocity dispersion is 0.3, which is not a particularly large value 
among clusters (see fig.~2 of Coziol et al. \cite{coz09}).  Moreover,
RX J105453.3+552102 seems well isolated in the phase-space as show by
Fig.~\ref{figvd} (see den Hartog \& Katgert \cite{den96}
and Aguerri et al. \cite{aguerri07} for other clusters). This supports the idea that RX J105453.3+552102 is far from an
  important  accretion episode. Nevertheless, we have shown that,
    albeit the overall velocity distribution of the spectroscopical
    galaxy sample belonging to RX J105453.3+552102 is Gaussian, there
    is a significant departure from Gaussianity in the outer regions
    ($R>2'$) which we have interpreted as a possible signature of
    radial anisotropy of the galaxies in the group outskirts. If
    confirmed in more detailed dynamical analysis, this will represent
    an interesting piece of information to be added into the fossil
    group information scenarios (D'Onghia et al. 2005; Sommer-Larsen
    2006).

The second signature of an old and undisturbed cluster comes from the BGG
itself. In fact, RX J105453.3+552102 BGG  shows a small $n$ S\'ersic index value ($n\approx2$) and clear discy isophotes in the external regions. According to the
findings of numerical simulations, surface brightness profiles with small $n$ values result from gas
rich mergers (e.g., Khochfar \& Burkert \cite{kho05}). If the RX J105453.3+552102 BGG
has indeed been formed from the merger of all major galaxies within
the inner regions of the system in very early times, then some of
these mergers would have been gas--rich. In contrast, the merger
between two clusters and the following equal--mass dry mergers of the
corresponding dominant ellipticals would produce surface brightness profiles with $n\approx4$. Another piece of evidence in favour of the above scenario
is the absence of multiple nuclei both from our photometric data
 and from spectroscopic data. I fact, we took three spectra with
different slit position angles crossing the BGG nucleus and giving equal $z$ values. 

The third, possible piece of evidence that RX J105453.3+552102
is a relaxed old cluster comes from the dip observed in the photometric LF at
$M_{r}\sim-19.5$. This dip is more prominent in the LF$_{\rm phot}$
computed with the galaxies located within 0.5 Mpc radius. Similar
bimodality in the LF has also been reported in other relaxed clusters
(Yagi et al. \cite{yagi02}) and groups (Hunsberger et
al. \cite{huns98}, Miles et al. \cite{mil04}, Mendes de Oliveira et
al. \cite{mendes06}).

We can conclude that although RX J105453.3+552102  do not follow the prescription of a fossil system, is a relaxed and old galaxy cluster with no indication of recent infall of $L^{*}$ galaxies and therefore, it is a genuine FG.

\subsection{RX J105453.3+552102 a relaxed and massive system at z$\approx 0.5$}

Our analysis indicates that RX J105453.3+552102 is a very massive
galaxy cluster already relaxed at z$\approx$0.5. This means that
$\approx 6$ Gyr ago this cluster was as massive as the Coma cluster
but more dynamically evolved. In a hierarchical structure formation
scenario, these very massive, relaxed systems are likely formed at low
redshift.  In order to understand how common is a cluster like RX
J105453.3+552102, we have searched in the Millennium Simulation
(Springel et al. \cite{springel05}; Boylan-Kolchin et
al. \cite{boylan09}) for halos with masses M$_{200}>1\times10^{15}
h_{70}^{-1}$ M$_{\odot}$ located at z$=0.5$. The total number of such
halos was only 9 within the volume covered by the Millennium
Simulation (having a box of comoving size of 500 $h_{100}^{-1}$
Mpc). We have also computed the magnitude difference between the brightest and
the second brightest galaxies located in each of those halos: only 3/9
have $\Delta m_{12}>2$. However, the Millennium Simulation has been
carried out using a normalization of the power spectrum with
$\sigma_8=0.9$. Using instead a lower normalization, $\sigma_8=0.8$,
in agreement with the most recent CMB and large-scale structure
analysis (e.g. Komatsu et al. 2010, and references therein), the
number density of such massive clusters at $z=0.5$ drops by about a
factor of three. Thus we expect the number of ``fossil'' clusters at
least as massive as RX J105453.3+552102 within the Millennium
Simulation volume at $z=0.5$ to be of order unity. Clearly, 
to decide whether the detection of a relaxed massive cluster like RX
J105453.3+552102 should be considered as a rare event for a standard
$\Lambda$CDM cosmology, one has to know the volume within which this
cluster has been found, i.e. the selection function of the
corresponding survey. We postpone the discussion of this issue to a
future analysis, which will also be based on a larger statistics of
fossil groups/clusters.

\begin{acknowledgements}
We would like to thank to the anonymous referee for useful comments. This article is based on observations made with the Nordic Optical Telescope and Telescopio Nazionale Galileo operated on the island of La Palma, in the Spanish Observatorio del Roque de los Muchachos of the Instituto de Astrof\'{\i}sica de Canarias. JALA and JMA were supported by the projects AYA2010-21887-C04-04 and by the Consolideer-Ingenio 2010 Program grant CSD2006-00070. EMC was supported from Padua University through grants 60A02-1283/10 and CPDA089220 and by Ministry of Education, University and Research (MIUR) through grant EARA 2004-2006.
\end{acknowledgements}

\longtab{1}{
\begin{longtable}{cccccccccc}
\caption{Photometric and kinematic properties of the galaxies observed with DOLORES} \\
\hline
\hline
SDSS Name & cz & member & comments & m$_{r, our}$ & $m_{u,SDSS}$ & $m_{g,SDSS}$ & $m_{r,SDSS}$ & $m_{i,SDSS}$ & $m_{z,SDSS}$\\
   & (km s$^{-1}$)   & & & & & & & &  \\
\hline
\endfirsthead
\caption{continued.}\\
\hline\hline
SDSS Name  & cz & member & comments & m$_{r, our}$& $m_{u,SDSS}$ & $m_{g,SDSS}$ & $m_{r,SDSS}$ & $m_{i,SDSS}$ & $m_{z,SDSS}$\\
   & (km s$^{-1}$)   & & & & & & & &  \\
\hline
\endhead
\hline
\endfoot
 SDSS J105425.15+552103.0 & 141485 $\pm$ 100 & 1 & E &        &   24.585 &  22.081 &  21.063 &  20.447 &  20.566 \\
 SDSS J105427.11+552247.4 & 139639 $\pm$ 108 & 1 & NE&         &  23.500 &  23.080 &  21.590 &  20.760 &  20.060 \\
 SDSS J105428.19+552058.0 & 191053 $\pm$ 130 & 0 & NE&         &  24.220 &  22.510 &  21.960 &  20.810 &  20.480 \\
 SDSS J105428.47+552123.4 & 139112 $\pm$  98 & 1 & NE&         &  21.850 &  22.660 &  21.750 &  21.190 &  20.960 \\
 SDSS J105429.56+551955.6 & 150572 $\pm$  72 & 0 & NE&         &  21.520 &  21.330 &  20.370 &  19.830 &  19.420 \\
 SDSS J105430.42+552116.5 & 141276 $\pm$  56 & 1 & NE&         &  24.700 &  22.820 &  21.370 &  20.290 &  19.490 \\
 SDSS J105431.81+552224.2 & 139446 $\pm$  57 & 1 & NE&         &  25.930 &  23.320 &  21.430 &  20.630 &  20.300 \\
 SDSS J105433.07+552345.3 & 118229 $\pm$ 116 & 0 & NE&         &  23.580 &  22.490 &  20.650 &  19.980 &  19.570 \\
 SDSS J105433.47+552153.9 & 139997 $\pm$ 134 & 1 & NE&         &  25.690 &  23.710 &  21.920 &  21.680 &  20.980 \\
 SDSS J105433.59+552315.0 & 139255 $\pm$ 100 & 1 & E &        &  22.140 &  21.580 &  20.660 &  20.330 &  20.190 \\
 SDSS J105434.56+552233.6 & 138980 $\pm$  96 & 1 & NE&         &  25.050 &  22.750 &  21.240 &  20.420 &  19.890 \\
 SDSS J105434.76+552052.0 & 139734 $\pm$  80 & 1 & NE&  20.91  &  23.060 &  22.610 &  20.850 &  20.050 &  19.760 \\
 SDSS J105435.18+552057.1 & 139862 $\pm$ 148 & 1 & NE&  20.62  &  26.140 &  22.500 &  20.680 &  19.910 &  19.460 \\
 SDSS J105435.48+551849.1 & 108230 $\pm$ 162 & 0 & NE&  20.84  &  22.150 &  21.750 &  20.630 &  20.520 &  20.500 \\
 SDSS J105438.24+552217.1 & 140766 $\pm$  69 & 1 & NE&  21.32  &  24.930 &  23.150 &  21.410 &  20.690 &  20.070 \\
 SDSS J105438.39+551806.2 & 115193 $\pm$ 102 & 0 & NE&         &  23.640 &  22.540 &  21.500 &  21.270 &  21.010 \\
 SDSS J105438.83+552019.8 & 141886 $\pm$  81 & 1 & E &  20.83  &  22.360 &  21.610 &  20.980 &  20.520 &  21.000 \\
 SDSS J105438.84+552311.4 & 139216 $\pm$ 100 & 1 & E &  21.46 &  23.030 &  22.780 &  21.710 &  21.310 &  21.420 \\
 SDSS J105439.21+552108.2 & 139441 $\pm$ 124 & 1 & NE&  21.52  &  26.600 &  23.460 &  21.410 &  20.780 &  20.450 \\
 SDSS J105439.78+552155.3 & 142896 $\pm$ 106 & 1 & NE&  20.49  &  25.300 &  22.000 &  20.840 &  20.080 &  19.750 \\
 SDSS J105440.08+551840.2 & 114765 $\pm$  58 & 0 & NE&  19.06  &  21.230 &  20.470 &  19.350 &  18.900 &  18.420 \\
 SDSS J105440.09+552056.2 & 139605 $\pm$ 110 & 1 & NE&  21.30  &  24.760 &  23.280 &  21.830 &  21.000 &  20.820 \\
 SDSS J105440.50+552316.0 &  72992 $\pm$ 100 & 0 & E &  21.59 &  22.780 &  22.210 &  21.810 &  21.590 &  21.520 \\
 SDSS J105440.71+552157.6 & 143384 $\pm$  58 & 1 & NE&  20.44  &  25.720 &  22.210 &  20.600 &  19.790 &  19.230 \\
 SDSS J105442.08+551903.3 & 137694 $\pm$  66 & 1 & NE&  20.93  &  23.870 &  22.340 &  20.960 &  20.200 &  19.840 \\
 SDSS J105442.56+552022.8 & 126728 $\pm$ 166 & 0 & NE&  21.25  &  22.860 &  22.150 &  21.410 &  20.960 &  20.550 \\
 SDSS J105442.81+551922.8 & 139712 $\pm$  78 & 1 & NE&  21.05  &  25.170 &  22.690 &  21.170 &  20.270 &  19.760 \\
 SDSS J105442.94+552008.4 & 138861 $\pm$  66 & 1 & NE&  20.38  &  25.850 &  22.660 &  20.740 &  19.810 &  19.520 \\
 SDSS J105443.61+552026.8 & 140214 $\pm$  40 & 1 & NE&  20.69  &  22.570 &  22.490 &  20.850 &  19.980 &  19.510 \\
 SDSS J105444.08+552059.4 & 140503 $\pm$  43 & 1 & NE&  20.00  &  26.290 &  21.850 &  20.360 &  19.480 &  19.100 \\
 SDSS J105444.19+552311.9 &  93821 $\pm$ 120 & 0 & NE&  20.78  &  23.020 &  21.800 &  21.010 &  20.650 &  20.520 \\
 SDSS J105444.90+552211.2 & 138104 $\pm$  51 & 1 & NE&  20.99  &  23.280 &  22.650 &  21.210 &  20.530 &  20.380 \\
 SDSS J105445.14+552044.4 & 139763 $\pm$  43 & 1 & NE&  20.02  &  21.880 &  21.370 &  20.100 &  19.450 &  19.120 \\
 SDSS J105445.26+552422.2 & 125001 $\pm$ 164 & 0 & NE&         &  22.620 &  22.150 &  21.260 &  20.790 &  20.630 \\
 SDSS J105445.51+552005.6 & 140125 $\pm$  84 & 1 & NE&  21.36  &  24.730 &  23.180 &  21.550 &  20.750 &  20.290 \\
 SDSS J105445.53+552233.3 & 137294 $\pm$ 100 & 1 & E &  21.57 &  23.650 &  21.970 &  20.290 &  19.430 &  19.050 \\
 SDSS J105446.25+552231.8 & 138849 $\pm$  70 & 1 & NE&  20.09  &  23.650 &  21.970 &  20.290 &  19.430 &  19.050 \\
 SDSS J105446.34+551919.6 & 129031 $\pm$ 100 & 0 & E &  21.54 &  23.350 &  22.880 &  21.650 &  21.130 &  20.760 \\
 SDSS J105446.60+552400.7 & 139962 $\pm$  58 & 1 & NE&  21.35  &  24.710 &  23.280 &  21.390 &  20.400 &  19.960 \\
 SDSS J105446.82+552104.1 & 139224 $\pm$  82 & 1 & NE&  21.26  &  23.040 &  23.070 &  21.430 &  20.510 &  20.310 \\
 SDSS J105447.04+551825.6 & 129392 $\pm$  80 & 0 & NE&         &  24.850 &  22.920 &  20.950 &  20.460 &  19.890 \\
 SDSS J105447.26+552340.4 & 139079 $\pm$  92 & 1 & NE&  20.17  &  26.700 &  22.080 &  20.420 &  19.530 &  19.280 \\
 SDSS J105447.65+552140.4 & 140759 $\pm$  94 & 1 & NE&  21.50  &  24.300 &  23.160 &  21.770 &  20.910 &  20.420 \\
 SDSS J105447.87+552111.4 & 136842 $\pm$  70 & 1 & NE&  21.39  &  23.480 &  22.860 &  21.650 &  20.880 &  20.360 \\
 SDSS J105448.05+552105.8 & 139333 $\pm$ 110 & 1 & NE&  20.87  &  26.270 &  23.230 &  21.210 &  20.180 &  20.010 \\
 SDSS J105448.49+552106.3 & 232784 $\pm$ 100 & 0 & E &  22.13 &  23.670 &  22.970 &  22.240 &  21.750 &  21.030 \\
 SDSS J105448.29+552129.1 & 140102 $\pm$  68 & 1 & NE&  21.09  &  24.710 &  23.370 &  21.440 &  20.380 &  20.300 \\
 SDSS J105448.78+551757.0 & 144715 $\pm$ 100 & 0 & E &        &  22.590 &  22.470 &  21.450 &  21.260 &  21.270 \\
 SDSS J105449.61+552148.0 & 142630 $\pm$  96 & 1 & NE&  21.55  &  23.440 &  23.610 &  21.580 &  20.700 &  20.090 \\
 SDSS J105449.69+551921.9 & 102091 $\pm$ 112 & 0 & NE&  18.94  &  21.760 &  20.510 &  19.150 &  18.590 &  18.230 \\
 SDSS J105449.71+552006.4 & 114798 $\pm$ 152 & 0 & NE&  21.61  &  25.430 &  23.070 &  21.780 &  21.050 &  20.920 \\
 SDSS J105450.31+552117.3 & 137712 $\pm$  70 & 1 & NE&  21.04  &  23.160 &  22.420 &  20.930 &  20.470 &  19.970 \\
 SDSS J105450.37+552229.2 & 139167 $\pm$  92 & 1 & NE&  20.55  &  24.430 &  22.400 &  20.620 &  19.720 &  19.300 \\
 SDSS J105451.15+551822.7 & 110305 $\pm$ 100 & 0 & E &        &  23.630 &  23.130 &  21.530 &  21.090 &  20.950 \\
 SDSS J105451.20+552205.5 & 137511 $\pm$ 128 & 1 & NE&  20.97  &  23.530 &  22.710 &  21.080 &  20.160 &  19.990 \\
 SDSS J105451.56+552036.4 & 138463 $\pm$ 182 & 1 & NE&  21.25  &  24.870 &  22.890 &  21.200 &  20.520 &  20.200 \\
 SDSS J105451.93+552058.8 & 140795 $\pm$  84 & 1 & NE&  21.46  &  24.470 &  23.200 &  21.620 &  21.560 &  20.560 \\
 SDSS J105452.17+552127.1 & 140812 $\pm$ 120 & 1 & NE&  21.56  &  25.770 &  23.360 &  21.570 &  20.630 &  20.140 \\
 SDSS J105452.03+552112.5 & 140223 $\pm$  39 & 1 & NE&  18.08  &  21.430 &  19.450 &  17.690 &  16.780 &  16.340 \\
 SDSS J105452.57+552111.6 & 137879 $\pm$  90 & 1 & NE&  21.42  &  25.110 &  23.210 &  22.190 &  21.190 &  20.910 \\
 SDSS J105452.80+552426.3 & 194294 $\pm$ 100 & 0 & E &        &  24.150 &  23.010 &  21.970 &  21.550 &  21.830 \\
 SDSS J105452.97+551732.7 & 115124 $\pm$ 100 & 0 & E &        &  23.350 &  21.910 &  21.070 &  20.700 &  20.240 \\
 SDSS J105453.22+551747.6 & 140330 $\pm$ 122 & 1 & NE&         &  25.960 &  23.090 &  21.200 &  20.260 &  19.900 \\
 SDSS J105453.25+551911.8 & 141601 $\pm$  52 & 1 & NE&  20.45  &  23.290 &  22.190 &  20.790 &  19.930 &  19.530 \\
 SDSS J105453.81+552122.6 & 141516 $\pm$  70 & 1 & NE&  20.88  &  25.020 &  23.030 &  21.110 &  20.060 &  19.710 \\
 SDSS J105453.92+552207.9 &  75671 $\pm$ 150 & 0 & NE&  21.26  &  22.540 &  22.890 &  21.350 &  21.000 &  20.840 \\
 SDSS J105454.00+552129.2 & 144198 $\pm$ 112 & 1 & NE&  21.51  &  25.740 &  23.220 &  21.370 &  20.710 &  20.090 \\
 SDSS J105454.06+552055.2 & 140987 $\pm$ 132 & 1 & NE&  21.47  &  25.630 &  23.330 &  21.380 &  20.430 &  20.050 \\
 SDSS J105454.37+552322.4 &  63639 $\pm$ 100 & 0 & E &  20.43 &  21.940 &  21.260 &  20.640 &  20.220 &  20.440 \\
 SDSS J105454.43+552332.2 & 139934 $\pm$ 108 & 1 & NE&  21.42  &  25.010 &  22.940 &  21.550 &  21.000 &  20.250 \\
 SDSS J105454.44+552256.2 & 140911 $\pm$  50 & 1 & NE&  21.09  &  25.000 &  22.870 &  20.870 &  20.230 &  19.770 \\
 SDSS J105454.52+552153.2 & 140719 $\pm$ 116 & 1 & NE&  20.56  &  21.630 &  21.510 &  20.790 &  20.570 &  20.220 \\
 SDSS J105454.57+552316.1 & 138810 $\pm$  88 & 1 & NE&  20.82  &  24.410 &  22.170 &  21.090 &  20.600 &  20.040 \\
 SDSS J105455.24+552348.7 & 140494 $\pm$  84 & 1 & NE&  20.56  &  24.120 &  22.740 &  20.830 &  19.920 &  19.410 \\
 SDSS J105455.24+552229.6 & 139666 $\pm$  96 & 1 & NE&  21.40  &  24.140 &  22.860 &  21.600 &  20.910 &  20.290 \\
 SDSS J105455.35+552106.4 & 139288 $\pm$  72 & 1 & NE&  20.55  &  23.070 &  22.410 &  20.630 &  19.750 &  19.350 \\
 SDSS J105456.23+552132.3 &  54707 $\pm$ 100 & 0 & E &  20.19 &  22.870 &  20.970 &  20.440 &  20.220 &  20.360 \\
 SDSS J105456.24+552119.5 & 138061 $\pm$  53 & 1 & NE&  20.94  &  25.630 &  22.860 &  21.300 &  20.350 &  19.970 \\
 SDSS J105456.36+551938.0 &  97145 $\pm$ 100 & 0 & E &  21.76 &  22.310 &  22.920 &  21.990 &  22.070 &  21.280 \\
 SDSS J105456.47+552106.1 & 138140 $\pm$ 180 & 1 & NE&  21.18  &  24.860 &  22.730 &  21.600 &  20.820 &  20.390 \\
 SDSS J105456.63+552252.9 & 137620 $\pm$  97 & 1 & NE&  20.48  &  25.940 &  22.450 &  20.550 &  19.820 &  19.370 \\
 SDSS J105456.65+551850.3 & 115040 $\pm$  74 & 0 & NE&  20.37  &  25.200 &  22.260 &  20.590 &  20.000 &  19.600 \\
 SDSS J105456.67+552056.2 & 140402 $\pm$ 112 & 1 & NE&  21.43  &  22.840 &  23.030 &  21.820 &  20.900 &  20.520 \\
 SDSS J105456.70+552100.2 & 140377 $\pm$  86 & 1 & NE&  21.25  &  25.780 &  22.950 &  21.540 &  20.740 &  20.250 \\
 SDSS J105457.29+552132.0 & 225652 $\pm$  71 & 0 & E &  22.18  &  24.970 &  22.570 &  21.580 &  20.860 &  20.640 \\
 SDSS J105458.04+552004.3 &  75840 $\pm$ 100 & 0 & E &  21.02 &  23.350 &  21.770 &  20.980 &  20.910 &  21.540 \\
 SDSS J105459.02+552059.6 & 141759 $\pm$  79 & 1 & NE&  21.84  &  24.570 &  23.140 &  21.980 &  21.070 &  21.120 \\
 SDSS J105459.80+551943.5 &  98567 $\pm$ 120 & 0 & NE&  19.54  &  21.540 &  20.710 &  19.750 &  19.320 &  19.140 \\
 SDSS J105502.36+552222.5 & 138289 $\pm$  66 & 1 & NE&  20.40  &  22.720 &  22.190 &  20.760 &  19.930 &  19.530 \\
 SDSS J105502.56+552428.4 & 145237 $\pm$ 100 & 0 & E &        &  22.570 &  22.710 &  21.540 &  20.950 &  20.760 \\
 SDSS J105502.82+552301.6 &  93992 $\pm$  66 & 0 & NE&  19.93  &  23.220 &  21.350 &  20.060 &  19.440 &  19.020 \\
 SDSS J105502.88+552145.9 & 185676 $\pm$ 100 & 0 & E &  20.17 &  22.150 &  21.330 &  20.340 &  19.540 &  19.240 \\
 SDSS J105503.17+551941.3 & 147211 $\pm$  98 & 0 & NE&  20.92  &  23.660 &  22.430 &  20.970 &  20.240 &  19.830 \\
 SDSS J105503.87+552154.0 & 142410 $\pm$  82 & 1 & NE&  20.52  &  24.330 &  22.550 &  20.920 &  20.140 &  19.700 \\
 SDSS J105503.87+552108.5 & 139865 $\pm$  60 & 1 & NE&  20.99  &  24.640 &  22.980 &  21.320 &  20.530 &  20.040 \\
 SDSS J105504.35+551923.6 & 146988 $\pm$ 100 & 0 & E &  21.39 &  22.590 &  22.620 &  21.880 &  21.370 &  21.290 \\
 SDSS J105505.49+552034.1 & 141445 $\pm$  98 & 1 & NE&  21.10  &  26.610 &  22.580 &  21.290 &  20.680 &  19.880 \\
 SDSS J105505.57+552245.3 & 139111 $\pm$ 122 & 1 & NE&  20.91  &  23.230 &  22.130 &  20.900 &  20.550 &  20.070 \\
 SDSS J105506.15+551914.6 & 191048 $\pm$ 100 & 0 & E &  21.53 &  22.800 &  22.510 &  21.650 &  21.490 &  21.430 \\
 SDSS J105506.18+551858.4 & 137366 $\pm$ 100 & 1 & E &  20.04 &  22.220 &  21.300 &  20.260 &  19.730 &  19.280 \\
 SDSS J105506.23+552257.7 & 126848 $\pm$  71 & 0 & E &  21.85 &  23.730 &  22.760 &  21.980 &  22.100 &  21.200 \\
 SDSS J105506.84+552306.3 & 140352 $\pm$ 138 & 1 & NE&  22.15  &  24.770 &  23.460 &  21.960 &  21.410 &  20.630 \\
 SDSS J105508.66+552121.8 & 141913 $\pm$  88 & 1 & NE&  20.78  &  24.170 &  22.780 &  21.100 &  20.440 &  20.190 \\
 SDSS J105509.72+552102.0 & 107957 $\pm$  59 & 0 & NE&  20.73  &  22.520 &  22.060 &  20.970 &  20.370 &  20.020 \\
 SDSS J105510.66+552253.3 & 108052 $\pm$ 100 & 0 & NE&  20.44  &  23.200 &  22.110 &  20.690 &  20.090 &  19.740 \\
 SDSS J105510.89+552323.3 & 138834 $\pm$ 166 & 1 & NE&  21.08  &  26.460 &  23.090 &  21.160 &  20.260 &  20.000 \\
 SDSS J105511.18+552255.8 &  75742 $\pm$ 100 & 0 & E &        &  22.810 &  21.570 &  20.970 &  20.640 &  20.260 \\
 SDSS J105511.57+552100.4 & 140472 $\pm$  96 & 1 & NE&         &  22.600 &  21.570 &  20.790 &  20.490 &  20.220 \\
 SDSS J105513.08+552315.2 & 138920 $\pm$  68 & 1 & NE&         &  24.760 &  22.630 &  20.810 &  20.010 &  19.400 \\
 SDSS J105513.63+552103.5 & 139550 $\pm$  84 & 1 & NE&         &  23.940 &  22.930 &  21.330 &  20.570 &  20.130 \\
 SDSS J105514.05+551952.9 & 190991 $\pm$  66 & 0 & NE&         &  22.010 &  21.420 &  20.460 &  19.840 &  19.610 \\
 SDSS J105515.13+552058.8 & 139317 $\pm$ 108 & 1 & NE&         &  23.470 &  22.990 &  21.970 &  21.320 &  21.040 \\
 SDSS J105515.94+552058.7 & 138504 $\pm$ 156 & 1 & NE&         &  22.850 &  23.190 &  21.630 &  21.000 &  20.300 \\
 SDSS J105516.06+551945.9 & 139489 $\pm$  84 & 1 & NE&         &  22.510 &  21.850 &  20.970 &  20.500 &  20.150 \\
 SDSS J105516.15+552030.8 & 140092 $\pm$ 150 & 1 & NE&         &  24.500 &  22.680 &  21.090 &  20.230 &  19.870 \\
 SDSS J105517.30+552108.0 & 139511 $\pm$  70 & 1 & NE&         &  22.280 &  20.840 &  19.150 &  18.340 &  17.880 \\
\label{catalogFG10}
\end{longtable}
}


\begin{thebibliography}{}

\bibitem[2007]{aguerri07} Aguerri, J.~A.~L., S{\'a}nchez-Janssen, R., \& Mu{\~n}oz-Tu{\~n}{\'o}n, C.\ 2007, \aap, 471, 17 

\bibitem[2006]{aguerri06} Aguerri, J.~A.~L., Castro-Rodr{\'{\i}}guez, N., Napolitano, N., Arnaboldi, M., \& Gerhard, O.\ 2006, \aap, 457, 771 


\bibitem[2005]{aguerri05} Aguerri, J.~A.~L., 
Gerhard, O.~E., Arnaboldi, M., Napolitano, N.~R., Castro-Rodriguez, N., 
\& Freeman, K.~C.\ 2005, \aj, 129, 2585 

\bibitem[2004]{aguerri04} Aguerri, J.~A.~L., 
Iglesias-Paramo, J., Vilchez, J.~M., 
\& Mu{\~n}oz-Tu{\~n}{\'o}n, C.\ 2004, \aj, 127, 1344

\bibitem[1994]{ash94} Ashman, K. M., Bird, C. M., \& Zepf, S. E. 1994, \aj, 108, 2348

\bibitem[1994]{bar94} Bardelli, S., Zucca, E., Vettolani, G., et al. 1994, \mnras, 267, 665

\bibitem[2009]{bar09} Barrena, R., Girardi, M., Boschin, W., \& Das{\'{\i}}, M.\ 2009, \aap, 503, 357 


\bibitem[1990]{bee90} Beers, T. C., Flynn, K., \& Gebhardt, K. 1990, \aj, 100, 32

\bibitem[1991]{bee91} Beers, T. C., Gebhardt, K.,
Forman, W., Huchra, J. P., \& Jones, C., 
 1991, \aj, 102, 1581

\bibitem[1992]{bee92} Beers, T. C., Gebhardt, K., Huchra, J. P., et al. 1992, \apj, 400, 410

\bibitem[1996]{ber96} Bertin, E., \& Arnouts, S. 1996, \aaps, 117, 393

\bibitem[1994]{bir94} Bird, C. M. 1994, \apj, 422, 480

\bibitem[2006]{biv06} Biviano, A., Murante, G., Borgani, S., Diaferio, A., Dolag, K., \& Girardi, M.\ 2006, \aap, 456, 23 

\bibitem[2004]{biv04} Biviano, A., \& Katgert, P. 2004, \aap, 424, 779

\bibitem[2000]{boe00} B\"ohringer, H., Voges, W., Huchra, J. P., et al.
2000, \apjs, 129, 435 

\bibitem[2009]{bos09} Boschin, W., Barrena, R., \& Girardi, M. 2009, \aap, 495, 15

\bibitem[2008]{bos08} Boschin, W., Barrena, R., Girardi, M., \& Spolaor, M. 2008, \aap, 487, 33

\bibitem[2009]{boylan09} Boylan-Kolchin, M., Springel, V., White, S.~D.~M., Jenkins, A., \& Lemson, G.\ 2009, \mnras, 398, 1150 

\bibitem[2004]{capak04} Capak, P., et al.\ 2004, 
\aj, 127, 180

\bibitem[1997]{car97} Carlberg, R. G., Yee, H. K. C., \& Ellingson, E. 1997, \apj, 478, 462

\bibitem[2009]{castror09} Castro-Rodrigu{\'e}z, N., Arnaboldi, M., Aguerri, J.~A.~L., Gerhard, O., Okamura, S., Yasuda, N., \& Freeman, K.~C.\ 2009, \aap, 507, 621 

\bibitem[2003]{castror03} Castro-Rodr{\'{\i}}guez, N., Aguerri, J.~A.~L., Arnaboldi, M., Gerhard, O., Freeman, K.~C., Napolitano, N.~R., \& Capaccioli, M.\ 2003, \aap, 405, 803 

\bibitem[1996]{colles96} Colless, M., \& Dunn, A.~M.\ 1996, \apj, 458, 435 

\bibitem[1976]{cou76} Cousins, A. W. J., 1976, Mem. R. Astr. Soc, 81, 25

\bibitem[2009]{coz09} Coziol, R., Andernacht, H., Caretta, C. a., Alamo-Martinez, K. A., Tago, E. 2009, \aj 137, 4795


\bibitem[2006]{cypriano06} Cypriano, E.~S., 
Mendes de Oliveira, C.~L., \& Sodr{\'e}, L., Jr.\ 2006, \aj, 132, 514 

\bibitem[1980]{dan80} Danese, L., De Zotti, C., \& di Tullio, G. 1980, \aap, 82, 322

\bibitem[2010]{dariush10} Dariush, A.~A., 
Raychaudhury, S., Ponman, T.~J., Khosroshahi, H.~G., Benson, A.~J., Bower, 
R.~G., \& Pearce, F.\ 2010, \mnras, 559 

\bibitem[1996]{den96} den Hartog, R., \& Katgert, P. 1996, \mnras, 279, 349

\bibitem[2005]{dimatteo05} Di Matteo, T., 
Springel, V., \& Hernquist, L.\ 2005, \nat, 433, 604 

\bibitem[2005]{donghia05} D'Onghia, E., 
Sommer-Larsen, J., Romeo, A.~D., Burkert, A., Pedersen, K., Portinari, L., 
\& Rasmussen, J.\ 2005, \apjl, 630, L109 

\bibitem[2004]{donghia04} D'Onghia, E., \& Lake, G.\ 2004, \apj, 612, 628

\bibitem[1988]{dre88} Dressler, A., \& Shectman, S. A. 1988, \aj, 95, 985

\bibitem[1994]{ell94} Ellingson, E., \& Yee, H. K. C. 1994, \apjs, 92, 33

\bibitem[1996]{fad96} Fadda, D., Girardi, M., Giuricin, G., Mardirossian, F., \& Mezzetti, M. 1996, \apj, 473, 670

\bibitem[1987]{fas87} Fasano, G., \& Franceschini, A. 1987, \mnras, 225, 155

\bibitem[1995]{fuk95} Fukugita, M., 
Shimasaku, K., \& Ichikawa, T.\ 1995, \pasp, 107, 945  

\bibitem[1991]{geb91} Gebhardt, K., \& Beers, T. C. 1991, \apj, 383, 72

\bibitem[1993]{gerhard93} Gerhard, O.~E.\ 1993, \mnras, 265, 213 

\bibitem[1996]{gir96} Girardi, M., Fadda, D., Giuricin, G. et al. 1996, \apj, 457, 61

\bibitem[1998]{gir98} Girardi, M., Giuricin, G., Mardirossian, F., Mezzetti, M., \& Boschin, W. 1998, \apj, 505, 74

\bibitem[2001]{gir01} Girardi, M., \& Mezzetti, M. 2001, \apj, 548, 79

\bibitem[2005]{gonzalez05} Gonzalez, A.~H., 
Zabludoff, A.~I., \& Zaritsky, D.\ 2005, \apj, 618, 195 

\bibitem[2002]{got02} Goto, T., Sekiguchi, M., Nichol, R. C., et al. 2002, \aj, 123, 1807

\bibitem[1996]{graham96} Graham, A., Lauer, 
T.~R., Colless, M., \& Postman, M.\ 1996, \apj, 465, 534 

\bibitem[2008]{hopkins08} Hopkins, P.~F., Hernquist, L., Cox, T.~J., Dutta, S.~N., \& Rothberg, B.\ 2008, \apj, 679, 156 

\bibitem[2001]{huang01} Huang, J.-S., et al.\ 2001, \aap, 368, 787 

\bibitem[1998]{huns98} Hunsberger, S.~D., 
Charlton, J.~C., \& Zaritsky, D.\ 1998, \apj, 505, 536

\bibitem[1987]{jedr87} Jedrzejewski, R.~I.\ 
1987, \mnras, 226, 747 

\bibitem[2003]{jones03} Jones, L.~R., Ponman, 
T.~J., Horton, A., Babul, A., Ebeling, H., 
\& Burke, D.~J.\ 2003, \mnras, 343, 627 

\bibitem[2000]{jones00} Jones, L.~R., Ponman, 
T.~J., \& Forbes, D.~A.\ 2000, \mnras, 312, 139 

\bibitem[1992]{ken92} Kennicutt, R. C. 1992, \apjs, 79, 225

\bibitem[2005]{kho05} Khochfar, S., \& Buckert, A. 2005, \mnras, 359, 1379

\bibitem[2007]{khosro07} Khosroshahi, H.~G., 
Ponman, T.~J., \& Jones, L.~R.\ 2007, \mnras, 377, 595 

\bibitem[2006]{khosro06} Khosroshahi, H.~G., 
Maughan, B.~J., Ponman, T.~J., \& Jones, L.~R.\ 2006, \mnras, 369, 1211 

\bibitem[2004]{khosro04}Khosroshahi, H.~G., 
Jones, L.~R., \& Ponman, T.~J.\ 2004, \mnras, 349, 1240 

\bibitem[Komatsu et al.(2010)]{2010arXiv1001.4538K} Komatsu, E., et al.\
2010, arXiv:1001.4538

\bibitem[2009]{labarbera09} La Barbera, F., de 
Carvalho, R.~R., de la Rosa, I.~G., Sorrentino, G., Gal, R.~R., 
\& Kohl-Moreira, J.~L.\ 2009, \aj, 137, 3942 

\bibitem[1960]{lim60} Limber, D. N., \& Mathews, W. G. 1960, \apj, 132, 286

\bibitem[2004]{lin04} Lin,Y.-T., \& Mohr, J. J. 2004, \apj, 617, 879

\bibitem[1992]{mal92} Malumuth, E. M., Kriss, G. A., Dixon, W. Van Dyke, Ferguson, H. C., \& Ritchie, C. 1992, \aj, 104, 495

\bibitem[2009]{mendes09} Mendes de 
Oliveira, C.~L., Cypriano, E.~S., Dupke, R.~A., 
\& Sodr{\'e}, L.\ 2009, \aj, 138, 502 

\bibitem[2006]{mendes06} Mendes de 
Oliveira, C.~L., Cypriano, E.~S., 
\& Sodr{\'e}, L., Jr.\ 2006, \aj, 131, 158

\bibitem[2008]{mendez08} M{\'e}ndez-Abreu, J., Aguerri, J.~A.~L., Corsini, E.~M., \& Simonneau, E.\ 2008, \aap, 478, 353

\bibitem[2001]{metcalfe01} Metcalfe, N., Shanks, 
T., Campos, A., McCracken, H.~J., \& Fong, R.\ 2001, \mnras, 323, 795 

\bibitem[2004]{mil04} Miles, T.~A., 
Raychaudhury, S., Forbes, D.~A., Goudfrooij, P., Ponman, T.~J., 
\& Kozhurina-Platais, V.\ 2004, \mnras, 355, 785 

\bibitem[2003]{mul03} Mullis, C. R., McNamara, B. R., Quintana, H, et al. 2003, \apj, 594, 154 


\bibitem[1986]{nag86} NAG Fortran Workstation Handbook, 1986 (Downers Grove, IL: Numerical Algorithms Group)

\bibitem[2002]{nelson02} Nelson, A.~E., Simard, 
L., Zaritsky, D., Dalcanton, J.~J., 
\& Gonzalez, A.~H.\ 2002, \apj, 567, 144 

\bibitem[2004]{ort04} Ortiz-Gil, A., Guzzo, L., Schuecker, P., B\"oringer, H.,
\& Collins, C. A. 2004, \mnras, 348, 325

\bibitem[2006]{patel06} Patel, P., Maddox, S., 
Pearce, F.~R., Arag{\'o}n-Salamanca, A., 
\& Conway, E.\ 2006, \mnras, 370, 851 


\bibitem[1996]{pin96} Pinkney, J., Roettiger, 
K., Burns, J.~O., \& Bird, C.~M.\ 1996, \apjs, 104, 1 

\bibitem[1993]{pis93} Pisani, A. 1993, \mnras, 265, 706

\bibitem[1996]{pis96} Pisani, A. 1996, \mnras, 278, 697

\bibitem[1997]{pog97} Poggianti, B.~M.\ 1997, \aaps, 122, 399 

\bibitem[1994]{ponman94} Ponman, T.~J., Allan, 
D.~J., Jones, L.~R., Merrifield, M., McHardy, I.~M., Lehto, H.~J., 
\& Luppino, G.~A.\ 1994, \nat, 369, 462 

\bibitem[2007a]{pop07a} Popesso, P., Biviano, A., B{\"o}hringer, H., \& Romaniello, M.\ 2007, \aap, 461, 397 

\bibitem[2007b]{pop07b} Popesso, P., Biviano, A., B{\"o}hringer, H., \& Romaniello, M.\ 2007, \aap, 464, 451 


\bibitem[1992]{pre92} Press, W. H., Teukolsky, S. A., Vetterling, W. T., \& Flannery, B. P. 1992, in Numerical Recipes (Second Edition), (Cambridge University Press)

\bibitem[2000]{qui00} Quintana, H., Carrasco, E. R., \& Reisenegger, A. 2000, \aj, 120, 511

\bibitem[2007]{ram07} Ramella, M., Biviano, A., Pisani, A. et
  al. 2007, \aap, 470, 39

\bibitem[2003]{rin03} Rines, K., Geller, M. J., Kurtz, M. J., \&
Diaferio, A. 2003, \aj, 126, 2152

\bibitem[2009]{roc09} Roche, N., Bernardi, M., 
\& Hyde, J.\ 2009, \mnras, 398, 1549

\bibitem[2009]{rudnick09} Rudnick, G., et al.\ 
2009, \apj, 700, 1559 

\bibitem[2005]{sanchez05} S{\'a}nchez-Janssen, R., Iglesias-P{\'a}ramo, J., Mu{\~n}oz-Tu{\~n}{\'o}n, C., Aguerri, J.~A.~L., \& V{\'{\i}}lchez, J.~M.\ 2005, \aap, 434, 521 

\bibitem[2007]{santos07} Santos, W.~A., Mendes de 
Oliveira, C., \& Sodr{\'e}, L., Jr.\ 2007, \aj, 134, 1551 

\bibitem[1976]{schechter76} Schechter, P.\ 1976, \apj, 
203, 297

\bibitem[2007]{seigar07} Seigar, M.~S., Graham, 
A.~W., \& Jerjen, H.\ 2007, \mnras, 378, 1575 

\bibitem[2006]{somer06} Sommer-Larsen, J.\ 2006, 
\mnras, 369, 958

\bibitem[2005]{springel05} Springel, V., et al.\ 2005, \nat, 435, 629 


\bibitem[2004]{sun04} Sun, M., Forman, W., 
Vikhlinin, A., Hornstrup, A., Jones, C., 
\& Murray, S.~S.\ 2004, \apj, 612, 805 

\bibitem[1986]{the86} The, L. S., \& White, S. D. M. 1986, \aj, 92, 1248

\bibitem[1979]{ton79} Tonry, J., \& Davis, M. 1979, \apj, 84, 1511

\bibitem[2010]{vandokkum10} van Dokkum, P.~G., 
et al.\ 2010, \apj, 709, 1018 

\bibitem[1998]{vikhlinin98} Vikhlinin, A., 
McNamara, B.~R., Forman, W., Jones, C., Quintana, H., 
\& Hornstrup, A.\ 1998, \apj, 502, 558 

\bibitem[2010]{vikram10} Vikram, V., Wadadekar, 
Y., Kembhavi, A.~K., \& Vijayagovindan, G.~V.\ 2010, \mnras, 401, L39 

\bibitem[2010]{voevodkin10} Voevodkin, A., 
Borozdin, K., Heitmann, K., Habib, S., Vikhlinin, A., Mescheryakov, A., 
Hornstrup, A., \& Burenin, R.\ 2010, \apj, 708, 1376 

\bibitem[2008]{vonbenda08} von 
Benda-Beckmann, A.~M., D'Onghia, E., Gottl{\"o}ber, S., Hoeft, M., 
Khalatyan, A., Klypin, A., {\ Muuml}ller, V.\ 2008, \mnras, 386, 2345 

\bibitem[1978]{wai78} Wainer, H., \&  Schacht, S. 1978, Psychometrika, 43, 203

\bibitem[1978]{white78} White, S.~D.~M., \& Rees, M.~J.\ 1978, \mnras, 183, 341

\bibitem[2002]{yagi02} Yagi, M., Kashikawa, N., 
Sekiguchi, M., Doi, M., Yasuda, N., Shimasaku, K., 
\& Okamura, S.\ 2002, \aj, 123, 87

\bibitem[2001]{yasuda01} Yasuda, N., et al.\ 
2001, \aj, 122, 1104 


\bibitem[2005]{zibetti05} Zibetti, S., White, 
S.~D.~M., Schneider, D.~P., \& Brinkmann, J.\ 2005, \mnras, 358, 949 

\end{thebibliography}
\end{document}